\documentclass[a4paper,11pt]{article}
\pdfoutput=1

\usepackage{tikz}
\usepackage{graphicx}
\usetikzlibrary{shapes.geometric, arrows}
\usepackage{multirow}

\usepackage[compat=1.0.0]{tikz-feynman}

\usepackage{jheppub}
\usepackage[T1]{fontenc} 
\usepackage{amsmath,amssymb}
\usepackage{ bbold }
\usepackage{braket} 
\usepackage{ mathrsfs }
\usepackage[small]{caption}
\usepackage{subcaption}
\usepackage{etoolbox,hyperref,makecell}
\usepackage{diagbox}
\usetikzlibrary{arrows.meta}
\usepackage{empheq}
\usepackage[freestanding]{hepunits}
\patchcmd{\abstract}{\null\vfil}{}{}{}
\newcommand{\bea}{\begin{eqnarray}}  
\newcommand{\eea}{\end{eqnarray}}

\usepackage{hhline}
\usepackage{float}
\usepackage{caption}

\usepackage[utf8]{inputenc}

\graphicspath{{./figures/}}

\title{On the Origin of Long-Lived Particles}

\author{Jared Barron and}
\author{David Curtin}

\affiliation{Department of Physics, University of Toronto, Canada}

\emailAdd{jared.barron@mail.utoronto.ca}
\emailAdd{dcurtin@physics.utoronto.ca}

\date{\today}

\abstract{
MATHUSLA is a proposed large-volume displaced vertex (DV) detector, situated on the surface above CMS and designed to search for long-lived particles (LLPs) produced at the HL-LHC.
We show that a discovery of LLPs at MATHUSLA would not only prove the existence of BSM physics, it would also uncover the theoretical origin of the LLPs, despite the fact that MATHUSLA gathers no energy or momentum information on the LLP decay products.
Our analysis is simple and robust, making it easily generalizable to include more complex LLP scenarios, and our methods are applicable to LLP decays discovered in ATLAS, CMS, LHCb, or other external detectors. 
In the event of an LLP detection, MATHUSLA can act as a Level-1 trigger for the main detector, guaranteeing that the LLP production event is read out at CMS.
We perform an LLP simplified model analysis to show that combining information from the MATHUSLA and CMS detectors would allow the LLP production mode topology to be determined with as few as $\sim 100$ observed LLP decays. 
Underlying theory parameters, like the LLP and parent particle masses, can also be measured with $\lesssim 10\%$ precision. 
Together with information on the LLP decay mode from the geometric properties of the observed DV, it is clear that MATHUSLA and CMS together will be able to characterize any newly discovered physics  in great detail.
}

\begin{document}

\maketitle

\newpage

\section{Introduction}
\label{section:intro}

Since the Higgs boson was discovered by the Large Hadron Collider (LHC) experiments ATLAS and CMS in 2012~\cite{Aad:2012tfa,Chatrchyan:2012xdj}, the Standard Model (SM) of particle physics has been complete, with every fundamental particle predicted having been found. Nonetheless, many questions remain unresolved, including the nature of the dark matter, neutrino masses, and baryogenesis, to name a few \cite{Ade:2015xua,1980ApJ...238..471R,Fukuda:1998mi,Ahmad:2001an}. All of these require the existence of physics beyond the Standard Model (BSM). In particular, symmetry-based solutions to the hierarchy problem of the Higgs boson require the existence of TeV-scale particles to avoid fine-tuning of the Higgs boson mass. Motivated by theories like supersymmetry (SUSY) \cite{Martin:1997ns} as well as composite or extra-dimensional models that can naturally give rise to such TeV-scale particles and solve the hierarchy problem \cite{Csaki:2018muy}, the efforts of the experimental particle physics community at the LHC have been largely devoted to searches for heavy, promptly decaying particles as the most likely avenue for discovery. The LHC main detectors are explicitly designed for such a search program. Unfortunately, despite strong theoretical motivations and a decade of searching, no BSM particles have been discovered to date~\cite{Aaboud:2017ayj,Aaboud:2018lpl,Sirunyan:2019glc,Aad:2019hjw}. 

As many canonical models to solve issues like the hierarchy problem are becoming highly constrained, 
it is important to consider other possible signatures of BSM physics that may have escaped notice so far. One of these possibilities is that BSM particles may exist at masses accessible by current colliders, but are long-lived rather than promptly decaying~\cite{Alimena:2019zri}. Particles with macroscopic decay lengths can elude detection at large collider experiments because of triggers designed for prompt decays, complicated backgrounds, and low acceptance due to a high probability of escaping the detector before decaying if the lifetime is long. 
Furthermore, LLPs are highly theoretically motivated in their own right, both from a bottom-up point of view, since various mechanisms resulting in meta-stable states in the SM can equally well operate in BSM theories, as well as from a top-down point of view, since LLPs arise and are instrumental in many BSM scenarios postulated to explain the hierarchy problem, dark matter, baryogenesis, and neutrino masses~\cite{Curtin:2018mvb}.

Recently, several new auxiliary detectors~\cite{Chou:2016lxi, Gligorov:2017nwh,Feng:2017uoz} and CERN experiments~\cite{Anelli:2015pba} have been proposed to extend the LLP reach of the existing LHC detectors. 
One of these is MATHUSLA (MAssive Timing Hodoscope for Ultra-Stable neutraL pArticles)~\cite{Chou:2016lxi,Alpigiani:2018fgd,Curtin:2018mvb}. 
MATHUSLA is a dedicated large-volume displaced vertex detector for the HL-LHC, to be positioned $\mathcal{O}(100)$ meters away from the LHC interaction point (IP). The detector will be composed of horizontal layers of trackers over an empty, air-filled decay volume. Its goal is to discover ultra-long lived particles produced at the LHC by reconstructing their decays inside its volume as displaced vertices (DVs). In the long lifetime regime, MATHUSLA can achieve cross-section sensitivity to weak-scale LLPs up to three orders of magnitude beyond ATLAS, depending on the decay and production mode \cite{Curtin:2018mvb}. This gain in sensitivity is largely due to MATHUSLA's much lower backgrounds for LLP signals than main detector experiments, made possible by copious shielding  from the LHC collision and the imposition of stringent signal reconstruction requirements. 

MATHUSLA is fundamentally a giant tracker without any energy or momentum measurement. 
This is well-suited to reconstructing DVs in a low-background environment, but it might not be obvious how to characterize any new particles that are discovered.
Fortunately, MATHUSLA is not only an LLP-discovery machine, and its physics utility goes beyond initial observation of LLP signals.  
Despite its simplicity, MATHUSLA can use geometric information to extract the likely velocity and decay mode of detected LLPs~\cite{Curtin:2017izq}. 
This also allows the LHC bunch crossing that produced the LLP to be identified. Furthermore, MATHUSLA can act as a Level-1 trigger for CMS, ensuring information about the LLP production event  is written to tape regardless of main detector trigger thresholds. 
Together, these abilities suggest that it should be possible to combine information from both MATHUSLA and CMS in order to study the properties of detected LLPs and details of the underlying theory. 

In this work we analyze the prospect of using correlated MATHUSLA and main detector information to diagnose the production mode of LLPs, and estimate underlying model parameters under a given production mode hypothesis. 
One initial obstruction to this effort is the vast space of possible models that give rise to LLPs. Fortunately, recent progress has been made to organize much of this space into a set of just a few `simplified models' that describe particular experimental signatures~\cite{Alimena:2019zri}. Complicated scenarios exist that do not fit easily into any of these categories, but the simplified model framework provides a good starting point for analyses that cover a wide range of commonly studied BSM scenarios in a relatively model-independent fashion. 

Assuming that MATHUSLA observes a number of LLP decays that all originate from a single production mode, we show that with simple cuts using only a few event observables, observation of $\sim \mathcal{O}(100)$ events allows the production mode to be determined with $\gtrsim 90\%$ confidence.  
Even for $\mathcal{O}(10)$ observed events, some simplified models can be correctly identified with high probability, while higher event yield obviously results in even greater classification confidence.
Assuming correct model classification, the LLP and parent particle masses can be determined to $\sim 10\%$ precision with similarly modest statistics.

 Our simple analysis can be generalized to include more production modes, and more complicated scenarios with several LLP production and decay modes considered simultaneously. Our methods should also be applicable to other proposed external LLP detectors~\cite{Gligorov:2017nwh,Feng:2017uoz}, as well as to LLPs discovered in the LHC main detectors themselves (where additional information on the LLP decay products may be available). 
 Together with information on the LLP decay mode from the geometric properties  of the DVs~\cite{Curtin:2017izq}, a discovery at MATHUSLA would therefore not only prove the existence of BSM physics, it would allow the theoretical origin of LLPs to be uncovered.

The paper is structured as follows. First we will review the MATHUSLA detector and its potential to be used as a trigger for CMS in Section \ref{section:detector}. 
In Section \ref{section:models} we will summarize the simulation of LLP production under several simplified models and across a wide range of parameters. 
In Section \ref{section:classification} we will develop a robust, physically motivated classifier to diagnose the LLP production mode using both CMS and MATHUSLA observables, and evaluate its performance. 
We describe how simple analyses can extract the LLP and parent particle masses for  the considered LLP production modes in  Section \ref{section:paramestimation}, and perform maximum likelihood estimation using pseudo-data compared to simulation-derived distributions to estimate the achievable precision. Section \ref{section:pileup} contains a brief discussion of complications due to multiple possible LLP production vertices, and we make concluding remarks in Section~\ref{section:conclusion}. 
Appendix~\ref{section:appendix} contains various Monte Carlo details for LLP event generation in our analysis.

\section{The MATHUSLA Detector}
\label{section:detector}
We first review the principles and geometry of the MATHUSLA detector in Section \ref{subsection:overview} to provide context for the rest of the analysis. Section \ref{subsection:physicsreach} contains a brief overview of the physics reach of MATHUSLA. In Section \ref{subsection:boost} we review how MATHUSLA can measure LLP boost, and the implications for coordinating between MATHUSLA and CMS.

	\subsection{Detector overview}
	\label{subsection:overview}
\begin{figure}
		\centering
		\includegraphics[width=.6\textwidth]{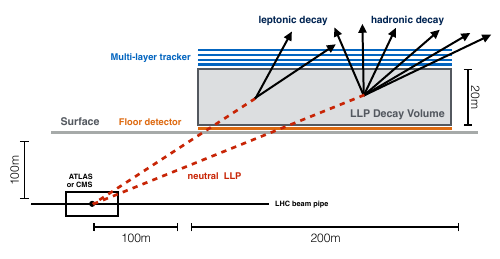}
		\caption{Schematic illustration of MATHUSLA, using the original 200~m benchmark geometry~\cite{Chou:2016lxi,Curtin:2018mvb}. Our results will apply to the more realistic geometries currently considered by the MATHUSLA collaboration, with $\sim (100 \ \mathrm{m})^2$ area situated closer to the collision point, as well. The tracking planes in the roof detect charged particles, allowing for the reconstruction of displaced vertices in the air-filled decay volume. The floor protector provides vetoing capability against charged particles entering the detector from below. 
		}
		\label{fig:MATHUSLA}
	\end{figure}

	The proposed MATHUSLA detector is a large-volume surface detector, to be placed near CMS \cite{Alpigiani:2018fgd,Lubatti:2019vkf}, see Figure~\ref{fig:MATHUSLA}. The goal is to instrument a large volume to search for displaced vertex (DV) decays of ultra long-lived particles produced at the LHC, focusing on decay lengths $c\tau \gtrsim 100 $m. It has been shown~\cite{Alpigiani:2018fgd} that MATHUSLA can operate in a low-background environment by making use of strong restrictions on the final state multiplicity, timing, and geometry of candidate events. The detector principle is to place several ($\sim$5) layers of trackers, likely to be plastic scintillator, above a large decay volume of air. LLPs decaying to SM charged particles inside MATHUSLA will give rise to upwards going tracks.
 By requiring reconstructed tracks to travel upwards relativistically and originate both spatially and temporally at a point inside the empty detector volume, backgrounds such as LHC muons, atmospheric neutrinos and cosmic rays  
 can be rejected. 
 
MATHUSLA will be equipped with an internal trigger system for LLPs, which relies on real-time tagging of upwards-traveling track-candidates within groups of neighboring detector modules~\cite{Alpigiani:2018fgd}. The trigger rate will be low enough that the MATHUSLA trigger can also act as a Level-1 burst-trigger for CMS: If the upwards tracks originate from the decay of an LLP, there is a range of $< 10$ candidate LHC bunch crossings that are very likely to include the production event at CMS. The MATHUSLA Level-1 trigger can therefore request a range of CMS events to be written to tape, which would enable the kind of off-line analysis we study in this work. If only CMS triggers were available, then the post-discovery analysis of many scenarios we consider here would be significantly more difficult or even impossible, due to much smaller trigger efficiencies in the main detector~\cite{Curtin:2018mvb}.

 In this work, we assume the original MATHUSLA benchmark geometry~\cite{Chou:2016lxi,Curtin:2018mvb}, which is a 200m $\times$ 200m square detector with a 20m high decay volume, placed 100m displaced from the interaction point both horizontally and vertically. More recently, the collaboration has considered more realistic designs with a 100m $\times$ 100m detector volume at an available site near CMS, with significantly smaller horizontal and vertical displacement compensating for the smaller size. This results in practically identical sensitivities to LLP decays~\cite{Alkhatib:2019eyo}. Our results apply almost verbatim to this more realistic design as well.

\subsection{Physics Reach}
	\label{subsection:physicsreach}
\begin{figure}
		\centering
		\includegraphics[width=.6\textwidth]{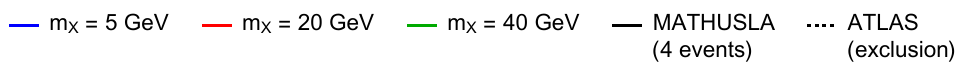}
		\\
		\includegraphics[width=.6\textwidth]{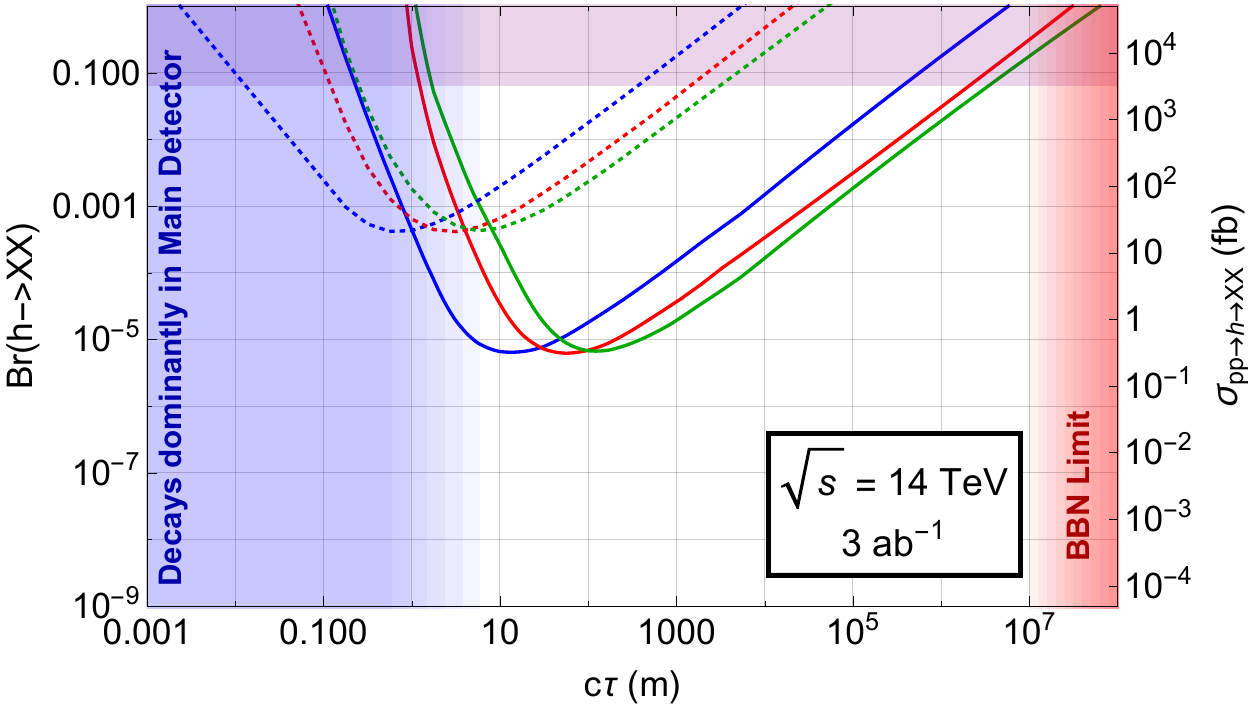}
		\caption{
		MATHUSLA sensitivity to new LLPs $X$ pair-produced in exotic Higgs decays, where $X$ decays hadronically. Exclusion curves correspond to 4 events in MATHUSLA.
		Plot from \protect{\cite{Curtin:2018mvb}}.
		}
		\label{fig:ExoticHiggsSensitivity}
	\end{figure}

	Due to MATHUSLA's near-zero backgrounds, $\sim 10$ observed events are likely enough for a discovery in most of the targeted LLP scenarios, especially if they decay to hadrons. This low threshold allows MATHUSLA to probe much lower cross-sections than ATLAS and CMS in the long-lifetime regime. The number of LLPs MATHUSLA will observe depends primarily on the production cross-section and decay length of the LLPs, and has only mild model dependence. In the long-lifetime regime, the lowest cross-section that MATHUSLA can exclude scales linearly with $\bar{b}c\tau$, where $\bar{b}$ is the mean boost of the LLPs. The minimum signal cross-section required for discovery is roughly~\cite{Curtin:2018mvb} 
	\begin{equation}
	\sigma_{\text{sig,min}}^{\text{LHC}} \sim (1\ \femtobarn)\left(\frac{\bar{b}c\tau}{10^{3}\ \meter}\right) \ .
	\end{equation}
	To give a concrete example of the detector's physics reach, Figure~\ref{fig:ExoticHiggsSensitivity} shows that MATHUSLA can probe exotic Higgs decay branching ratios to LLPs down to $10^{-5}$. For this study we will work in terms of the number of observed LLP decays, $N_{obs}$, ranging from 10 to 1000 without specifying the corresponding cross-sections for each model under consideration. Clearly this is a plausible event yield for many BSM scenarios.
	
	\subsection{Measuring LLP boost}
	\label{subsection:boost}

There are two important reasons to measure LLP boost at MATHUSLA~\cite{Curtin:2017izq}. First, the boost is highly correlated with LLP mass for a given production mode. Second, knowledge of LLP boost allows for the LHC production event to be identified, or at least narrowed down to a few possible bunch crossings, which allows us to use information from both MATHUSLA and CMS to learn about the nature of newly detected LLPs. 

Ref.~\cite{Curtin:2017izq} showed that MATHUSLA can determine the velocity of detected LLPs using only geometric information of a displaced vertex and the knowledge of LLP's direction of travel, from the LHC IP to the reconstructed DV. For a two-body decay, the velocity of the LLP decaying to two particles at angles $\theta_{1},\theta_{2}$ is 	
	\begin{equation}\beta_{X} = \frac{\beta_{1}\beta_{2}\sin(\theta_{1}+\theta_{2})}{\beta_{1}\sin(\theta_{1}) + \beta_{2}\sin(\theta_{2})}
	\end{equation}
	Since in the vast majority of cases the LLP decay products are ultra-relativistic, $\beta_{1}$ and $\beta_{2}$ can be set to 1 while incurring only negligible error. Regardless, the timing resolution of the proposed tracking detectors should allow these velocities to be determined within $\sim 5\%$. The uncertainty in $\theta_{1}$ and $\theta_{2}$ is dominated by detector effects and is estimated at  roughly $0.2\%$ for cm spatial resolution.
	
	In the case of an LLP decay mode whose final states have high multiplicity, i.e. hadronic decays, there is an alternative method to estimate its velocity, relying on the approximate sphericity of charged decay products in the LLP rest frame. Labelling the final-state momenta $p_{i}$, we solve for $\beta_{X}$ in the constraint 
	\begin{equation}
	\hat{p}_{X}\cdot \sum_{i} \hat{p}_{i}(\beta_{X}) = 0
	\end{equation}
	Essentially, the measured track momenta are boosted backward along the direction toward the LHC IP until they are approximately spherically distributed, which approximates the LLP rest frame. The accuracy of this estimate is negatively affected by the fact that the decay product momenta may not be lightlike as well as the fact that some decay products might be downgoing in the lab frame and therefore not included in the calculation. 
	Thus the measured LLP boost distribution suffers some systematic bias (which can be accounted for) and a small increase in spread.  Nonetheless, the LLP boost can be determined sufficiently well event-by-event to narrow down the candidate bunch crossings in which it was produced to $\sim 5$. 
There may be multiple hard-scattering vertices in the list of candidate events, but as discussed in \ref{section:pileup}, this does not jeopardize our analysis strategy.

\section{LLP Simplified Models}
	\label{section:models}
		
The space of all physical models that could produce long-lived particles is immense. To identify, consider, and discard them one by one would be enormously impractical. 
Simplified models break down BSM theories into smaller effective models that reflect experimental signatures and can be characterized by a small number of parameters that correspond to experimental observables, like cross sections, branching ratios, and masses.
In the past decade, the use of simplified models in searches for prompt particles at the LHC has become a standard practice \cite{Alves:2011wf}. 
Recently, considerable work has been done to extend the simplified model framework to LLPs~\cite{Alimena:2019zri}. This includes, in particular, separate simplified models for LLP production and decay, since they do not a priori have to be related. This is very suitable for our analysis, since the decay mode can be reasonably well characterized by MATHUSLA alone~\cite{Curtin:2017izq}, and we focus on the production mode.
In this section, we review and slightly expand upon the simplified model basis for LLP production modes introduced in~\cite{Alimena:2019zri}, and briefly explain how we generate Monte Carlo event samples for our study. 

\subsection{LLP Production Modes}

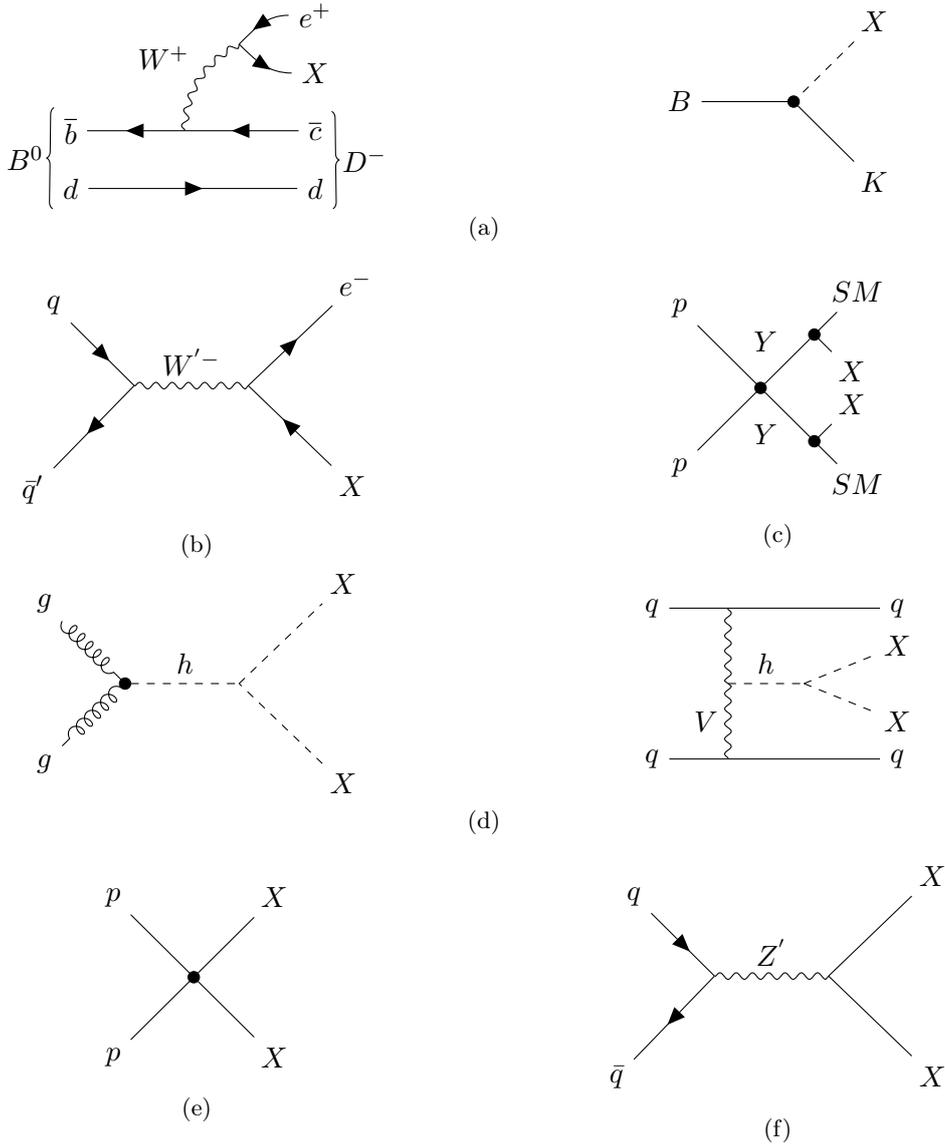
\begin{figure}
		\begin{subfigure}{.5\textwidth}
			\centering
			\begin{tikzpicture}
			\begin{feynman}
			\vertex (a1) {\(\overline b\)};
			\vertex[right=1.5cm of a1] (a2);
			\vertex[right=1.5cm of a2] (a3) {\(\overline c\)};
			\vertex[below=2em of a1] (b1) {\(d\)};
			\vertex[below=2em of a3] (b2) {\(d\)};
			\vertex[above=2em of a3] (c1) {\(X\)};
			\vertex[above=2em of c1] (c3) {\(e^{+}\)};
			\vertex at ($(c1)!0.5!(c3) - (1cm, 0)$) (c2);
			\diagram* {
				(a3) -- [fermion] (a2) -- [fermion] (a1),
				(b1) -- [fermion] (b2),
				(c3) -- [fermion, out=180, in=45] (c2) -- [fermion, out=-45, in=180] (c1),
				(a2) -- [boson, bend left, edge label=\(W^{+}\)] (c2),
			};
			\draw [decoration={brace}, decorate] (b1.south west) -- (a1.north west)
			node [pos=0.5, left] {\(B^{0}\)};
			\draw [decoration={brace}, decorate] (a3.north east) -- (b2.south east)
			node [pos=0.5, right] {\(D^{-}\)};
			\end{feynman}
			\end{tikzpicture}
		\end{subfigure}
		\begin{subfigure}{.5\textwidth}
			\centering
			\begin{tikzpicture}
			\begin{feynman}
			\vertex (i) {\(B\)};
			\vertex [dot, right= of i] (a){};
			\vertex [below right= of a] (b) {\(K\)};
			\vertex [above right= of a] (c) {\(X\)};
			\diagram* {
				(i) -- (a),
				(c) -- [scalar] (a) -- (b)
			};
			\end{feynman}
			\end{tikzpicture}				
		\end{subfigure}\\
		\begin{subfigure}{\textwidth}
		\caption{}
		\end{subfigure}\\
		\begin{subfigure}{.5\textwidth}
		\centering
		\begin{tikzpicture}
			\begin{feynman}
			\vertex (i1)  {\(q\)};
			\vertex [below right= of i1] (a1);
			\vertex [below left= of a1] (i2) {\(\bar{q}'\)};
			\vertex [right= of a1] (a2);
			\vertex [above right = of a2] (f1) {\(e^{-}\)};
			\vertex [below right = of a2] (f2) {\(X\)};
			\diagram* {
				(i1) -- [fermion] (a1) -- [fermion] (i2),
				(a1) -- [boson, edge label=\(W^{'-}\)] (a2),
				(f2) -- [fermion] (a2) -- [fermion] (f1)
			};
			\end{feynman}
		\end{tikzpicture}
		\caption{}
		\end{subfigure}
		\begin{subfigure}{.5\textwidth}
		\centering
			\begin{tikzpicture}
			\begin{feynman}
			\vertex (i1)  {\(p\)};
			\vertex [dot, below right= of i1] (a1){};
			\vertex [below left= of a1] (i2) {\(p\)};
			\vertex [dot, above right= 1cm of a1] (h1){};
			\vertex [dot, below right= 1cm of a1] (h2){};
			\vertex [above right = .8cm of h1] (f11) {\(SM\)};
			\vertex [below right = .7cm of h1] (f12) {\(X\)};
			\vertex [above right = .7cm of h2] (f21) {\(X\)};
			\vertex [below right = .8cm of h2] (f22) {\(SM\)};
			\diagram* {
				(i1) -- (a1) -- (i2),
				(a1) -- [edge label=\(Y\)](h1),
				(a1) -- [edge label'=\(Y\)](h2),
				(h1)  -- (f11),
				(h1) -- (f12),
				(h2) -- (f21),
				(h2) -- (f22),
			};
			\end{feynman}
			\end{tikzpicture}	
			\caption{}
		\end{subfigure}\\
		\begin{subfigure}{.5\textwidth}
			\centering
			\begin{tikzpicture}
			\begin{feynman}
			\vertex (i1)  {\(g\)};
			\vertex [dot, below right= of i1] (a1){};
			\vertex [below left= of a1] (i2) {\(g\)};
			\vertex [right= of a1] (a2);
			\vertex [above right = of a2] (f1) {\(X\)};
			\vertex [below right = of a2] (f2) {\(X\)};
			\diagram* {
				(i1) -- [gluon] (a1) -- [gluon] (i2),
				(a1) -- [scalar, edge label=\(h\)] (a2),
				(f2) -- [scalar] (a2) -- [scalar] (f1)
			};
			\end{feynman}
			\end{tikzpicture}	
		\end{subfigure}
		\begin{subfigure}{.5\textwidth}
			\centering
			\begin{tikzpicture}
			\begin{feynman}
			\vertex (i1)  {\(q\)};
			\vertex [below= 2cm of i1] (i2) {\(q\)};
			\vertex [right= 1cm of i1] (a1);
			\vertex [right= 1cm of i2] (a2);
			\vertex [below= 1cm of a1] (b1);
			\vertex [right= 1cm of b1] (b2);
			\vertex [right= 2cm of a1] (f1) {\(q\)};
			\vertex [right= 2cm of a2] (f4) {\(q\)};
			\vertex [below= .5cm of f1] (f2) {\(X\)};
			\vertex [below= 1cm of f2] (f3) {\(X\)};
			\diagram* {
				(i1) -- (a1) -- [boson] (b1),
				(i2) -- (a2) -- [boson, edge label = \(V\)] (b1),
				(b1) -- [scalar, edge label = \(h\)] (b2),
				(a1) -- (f1),
				(a2) -- (f4),
				(b2) -- [scalar] (f2),
				(b2) -- [scalar] (f3)			
			};
			\end{feynman}
			\end{tikzpicture}
		\end{subfigure}\\
		\begin{subfigure}{\textwidth}
			\caption{}
		\end{subfigure}\\
		\begin{subfigure}{.5\textwidth}
		\centering
		\begin{tikzpicture}
		\begin{feynman}
		\vertex (i1)  {\(p\)};
		\vertex [dot, below right= of i1] (a1){};
		\vertex [below left= of a1] (i2) {\(p\)};
		\vertex [above right = of a1] (f1) {\(X\)};
		\vertex [below right = of a1] (f2) {\(X\)};
		\diagram* {
			(i1) -- (a1) [dot] -- (i2),
			(f2) -- (a1) [dot] -- (f1)
		};
		\end{feynman}
		\end{tikzpicture}	
		\caption{}
		\end{subfigure}
		\begin{subfigure}{.5\textwidth}
		\centering
		\begin{tikzpicture}
		\begin{feynman}
		\vertex (i1)  {\(q\)};
		\vertex [below right= of i1] (a1);
		\vertex [below left= of a1] (i2) {\(\bar{q}\)};
		\vertex [right= of a1] (a2);
		\vertex [above right = of a2] (f1) {\(X\)};
		\vertex [below right = of a2] (f2) {\(X\)};
		\diagram* {
			(i1) -- [fermion] (a1) -- [fermion] (i2),
			(a1) -- [boson, edge label=\(Z^{'}\)] (a2),
			(f2) -- (a2) -- (f1)
		};
		\end{feynman}
		\end{tikzpicture}
		\caption{}
		\end{subfigure}
		\caption{Schematic Feynman diagrams for the models we consider. (a): Exotic $B$-meson decay (\textbf{BB}), with heavy neutral lepton and scalar LLPs. (b): Charged Current (\textbf{CC}). (c): Heavy Parent (\textbf{HP}). (d): Exotic Higgs Decay (\textbf{HIG}) with gluon fusion and vector boson fusion production channels. (e): Direct Pair Production (\textbf{DPP}). (f): Heavy Resonance (\textbf{RES}).} 
		\label{fig:feynmandiagrams}
	\end{figure}

In this paper we will employ the LLP simplified models recently defined and presented in the LHC LLP white paper \cite{Alimena:2019zri}, with the addition of LLP production in exotic $B$-meson decays. This set of simplified models, each representing an LLP production event topology, is meant to cover a wide range of well-motivated LLP production scenarios while remaining agnostic of the underlying mechanism generating the particles' long lifetimes. They are as follows, with schematic Feynman diagrams for each model in Figure~\ref{fig:feynmandiagrams}. 
	
	\begin{enumerate}
		\item The Exotic $B$-meson Decay channel is abbreviated as \textbf{BB} and has a the SM $B$-meson decay to an LLP plus one or more SM products. This scenario is not included in the simplified model library constructed in \cite{Alves:2011wf}, but we include this mode due to its importance for low-mass LLP benchmark models~\cite{Beacham:2019nyx} below $\sim 5$ GeV. 
		Two distinct scenarios for this production channel are right-handed neutrino models~\cite{Bondarenko:2018ptm,Chun:2019nwi} and light dark scalars \cite{Evans:2017lvd}. Both are explored in this paper.
	
		\item In the Charged Current (\textbf{CC}) scenario,  a BSM particle  possessing a SM electric charge is produced in the $s$-channel and decays to a charged SM particle and an LLP. In this case, assuming the dark sector particle is of integer charge, the charged SM decay product can be assumed to be a charged lepton. Models containing $W'$ bosons~\cite{Pati:1974yy,PhysRevD.11.2558,PhysRevD.12.1502} decaying to long-lived right-handed neutrinos~\cite{PhysRevLett.50.1427,Deppisch:2015qwa} are especially representative of this topology.

		\item In the Heavy Parent (\textbf{HP}) scenario some BSM particle is pair produced before each decays to an LLP and one or more standard model particles.
		The visible SM particles in this case could be quarks or leptons, leading to a signature in CMS of hard leptons or jets accompanying the observation of an LLP in MATHUSLA.\footnote{It is also possible, though less commonly considered, that the SM objects produced in association with the LLP are neutrinos. This challenging case of LLPs produced in association with exclusively invisible particles deserves further study.}
		Supersymmetry~\cite{Martin:1997ns} is full of examples that fall into this topology, since a dominant signature is superpartner pair production with subsequent decays to the lightest detector-stable supersymmetric state.
		HP scenarios with hard leptons or photons in the final state are relatively conspicuous and easy to diagnose, so we focus on jetty final states. Squark (gluino) pair production and subsequent decay (via an off-shell squark) to a long-lived neutralino would produce one (two) jet(s) per decay chain. 
	We consider both possibilities separately, and comment on how our methods could be easily extended to include leptonic or multi-step decay chains.

		\item In the Exotic Higgs Decay (\textbf{HIG}) scenario, the Standard Model Higgs boson couples to the dark sector and decays to two long-lived particles. Hidden sector models with a Higgs portal fall into this category \cite{Curtin:2013fra,Beacham:2019nyx}. More complex Higgs decay final states are of course a possibility, especially with decays into confining dark sectors \cite{Strassler:2006im, Strassler:2006ri, Craig:2014aea,Craig:2015pha}. The two-particle final state serves as a starting point when considering this space of models.
	
		\item In the Direct Pair Production (\textbf{DPP}) scenario, two long-lived particles are produced non-resonantly, through some effective operator, such as that generated by a very heavy squark leading to neutralino pair production, or pair-production of a hidden sector LLP via an off-shell $s$-channel mediator.

		\item The Heavy Resonance (\textbf{RES}) model has some BSM particle, such as a Z' boson (see e.g.~\cite{Yu:2013wta,Das:2019fee,Chiang:2019ajm}), produced on-shell in the s-channel that decays to two long-lived particles. In the high-mass limit, the Heavy Resonance model and Direct Pair Production model should coincide.
	\end{enumerate}

	We will use these simplified models as our `basis set' of LLP production topologies. Many searches at the LHC constrain the available parameter space for these models. However, in general, production cross-sections and branching ratios can be dialed down while still ensuring events in MATHUSLA, so we do not concern ourselves with current bounds on these signatures, except to note that they are not excluded for weak-scale masses.

\subsection{Event Generation}
	
	\label{subsection:eventgen}
	Event generation was performed with MadGraph5 version 2.6.6~\cite{Alwall:2014hca}, parton showering and hadronization  with Pythia8~\cite{Sjostrand:2006za,Sjostrand:2007gs}, and CMS detector simulation with~DELPHES 3.4.2 \cite{deFavereau:2013fsa}. For each of the simplified models, a corresponding simulation model was chosen from the LLP Simplified Model Library of~\cite{Alimena:2019zri} or existing standard MadGraph models \cite{Nilles:1983ge,Haber:1984rc,Rosiek:1989rs,Skands:2003cj,Allanach:2008qq,Martin:1997ns,PhysRevD.59.057504,PhysRevD.22.178,Kauffman1999}, and used to generate events across a range of mass-parameters relevant for HL-LHC production. 
	
	All of these simplified models have either one or two parameters that are varied in simulation. The only parameter that can be freely chosen in the BB, DPP, and HIG models is the LLP mass $m_{LLP}$. The CC, HP, and RES models also have a parent particle that decays to the LLP, so the two free parameters are $m_{parent}$ and $m_{LLP}$. The width of BSM parent particles can in principle be varied as well. For simplicity, $\Gamma_{parent}$ is set to $0.01\times m_{parent}$ for each of these models, since the exact width does not matter as long as it is small enough. A high-width parent particle case is also simulated for the RES model, with $\Gamma_{parent} = 0.3\times m_{parent}$, to illustrate the effect of parent particle width on classification accuracy.

	The range of LLP and parent masses for which simulation was generated in each model is summarized in Table \ref{tab:simparams}. 
	In single-parameter models, the chosen $m_{LLP}$  values are equally spaced within the indicated ranges (on a linear scale). 
	In models with a variable parent mass, $m_{parent}$ and $m_{LLP}/m_{parent}$ values are equally spaced within the indicated ranges. 
	This is relevant in interpreting our probabilistic results regarding the achievable accuracy in classifying datasets of observed LLP decays. 
	 Further details on the simulation of each simplified model are in appendix \ref{section:appendix}.

	\begin{table}
	\centering
	\begin{tabular}{|c|c|c|c|}
		\hline
		Model & MadGraph model & $m_{parent}$ range [GeV] & $m_{LLP}$ range \\
		\hline
		B decay & SM & $\sim 5.3$ GeV & $0.1 - 4$ GeV\\
		Charged Current & ChargeWN \cite{Alimena:2019zri} &  $50 - 1850$ GeV & $0.05 - 0.9 \times$ $m_{parent}$\\
		Heavy Parent & MSSM \cite{Nilles:1983ge,Haber:1984rc,Rosiek:1989rs,Skands:2003cj,Allanach:2008qq,Martin:1997ns,PhysRevD.59.057504} &  $50 -1850$ GeV & $0.05 - 0.9 \times$ $m_{parent}$\\ 
		Exotic Higgs Decay & heft \cite{PhysRevD.22.178,Kauffman1999}& 125 GeV&  $20 - 55$ GeV\\
		Direct Pair Production & MSSM & --& $50 - 2000$ GeV\\
		Heavy Resonance & $Zp_{2LLP}$ \cite{Alimena:2019zri} & $50 - 1850$ GeV& $0.05 - 0.4 \times$ $m_{parent}$\\
		\hline
	\end{tabular}
	\caption{Summary of signal simulation parameters. The values of $m_{LLP}$ or $(m_{parent}, m_{LLP}/m_{parent})$ used for simulation samples are equally spaced within the above indicated ranges on a linear scale.
} 
	\label{tab:simparams}	
	\end{table}

\subsection{LLP decay in MATHUSLA}
	For each point in parameter space, $10^{6}$ events were initially simulated. Only events with LLPs in the right rapidity range $\eta \in (0.8, 2)$ to potentially intersect MATHUSLA were considered. To increase the efficiency of our simulations for LLP decays in MATHUSLA we took advantage of the fact that the orientation of each event around the beam axis is random. Therefore, we rotated and appropriately reweighed events to assign LLPs a random azimuthal angle that guaranteed them flying through some part of MATHUSLA. 
	 Each event was also weighted to account for the probability of the LLP decaying inside MATHUSLA by a factor $L/bc\tau$, where $L$ is the distance each particular LLP travels through MATHUSLA, $b$ is its boost, and $c\tau$ is its decay length. In the long lifetime regime, this is proportional to the probability that the LLP decays within MATHUSLA. This event selection and weighting defines our high-statistics weighted samples. 
	 Since our analysis assumes some number of observed LLP decays $N_{obs}$, there is no weighting applied for a particular cross-section, and our results are valid for arbitrary $\bar b c\tau\gg 10^{2}\ \meter$, where $\bar b$ is average LLP boost. (Repeating our analysis for shorter lifetimes would be straightforward, except it would allow for additional information on the LLP lifetime to be extracted from the geometric distribution of decays in the detector.)
	 For each of these sets of weighted events, an unweighted event sample of size $\sim \mathcal{O}(10^4)$ was drawn. These are used to construct representative signal samples of $N_{obs} = 10, 100, 1000$ observed events. 
	 
	 	We make two  assumptions that simplify our analysis and simulations.
		 The first is that however the LLP decays, MATHUSLA will accurately measure its boost using the methods outlined in Section \ref{subsection:boost}. This allows us to omit detailed MATHUSLA detector simulation of the LLP reconstruction, and makes our analysis independent of details of the final detector design. 
	The second assumption is that each LLP decay can be uniquely matched to a hard scattering and hence a single primary vertex in an event at CMS. 
	This is to simplify our first pilot study of analyzing LLP events in MATHUSLA and CMS, and a more realistic analysis can be generalized to account for the possibility that each LLP decay is associated with several possible hard-scattering candidate events due to high pile-up at the HL-LHC. As we discuss in Section~\ref{section:pileup}, this complication should not qualitatively impact our conclusions on production mode classification and parameter estimation.

\section{Determining Production Mode}
\label{section:classification}

In this Section we will describe the strategy for identifying the LLP production mechanism at the LHC using combined data from CMS and MATHUSLA.

We assume that MATHUSLA observes $N_{obs} = 10, 100$ or 1000 LLP decays, all resulting from the same single production topology. 
 \emph{Sample-level variables} describing characteristics of the entire observed LLP dataset, like fraction of events with some number of jets above some $p_T$ in CMS, are used to classify the production mode. The algorithm is summarized in Figure~\ref{fig:Flowchart} and Table~\ref{tab:classificationalgorithm}.

Using characteristic features of each production mode, we find that simple cuts in sample-level observables can be used to achieve $\gtrsim 90\%$ probability of correct model classification for all but small corners of BSM particle parameter space with $\mathcal{O}(100)$ observed events, and $\gtrsim 98\%$ with $\mathcal{O}(1000)$ observed events. For the BB, CC, and HP models, $>90\%$ probabilities of correct classification can be achieved with only $N_{obs} = 10$ events.

\subsection{Hierarchical Classification Algorithm}
	\label{subsection:classmethod}

\begin{figure}
		\centering
		\includegraphics[width=.9\textwidth]{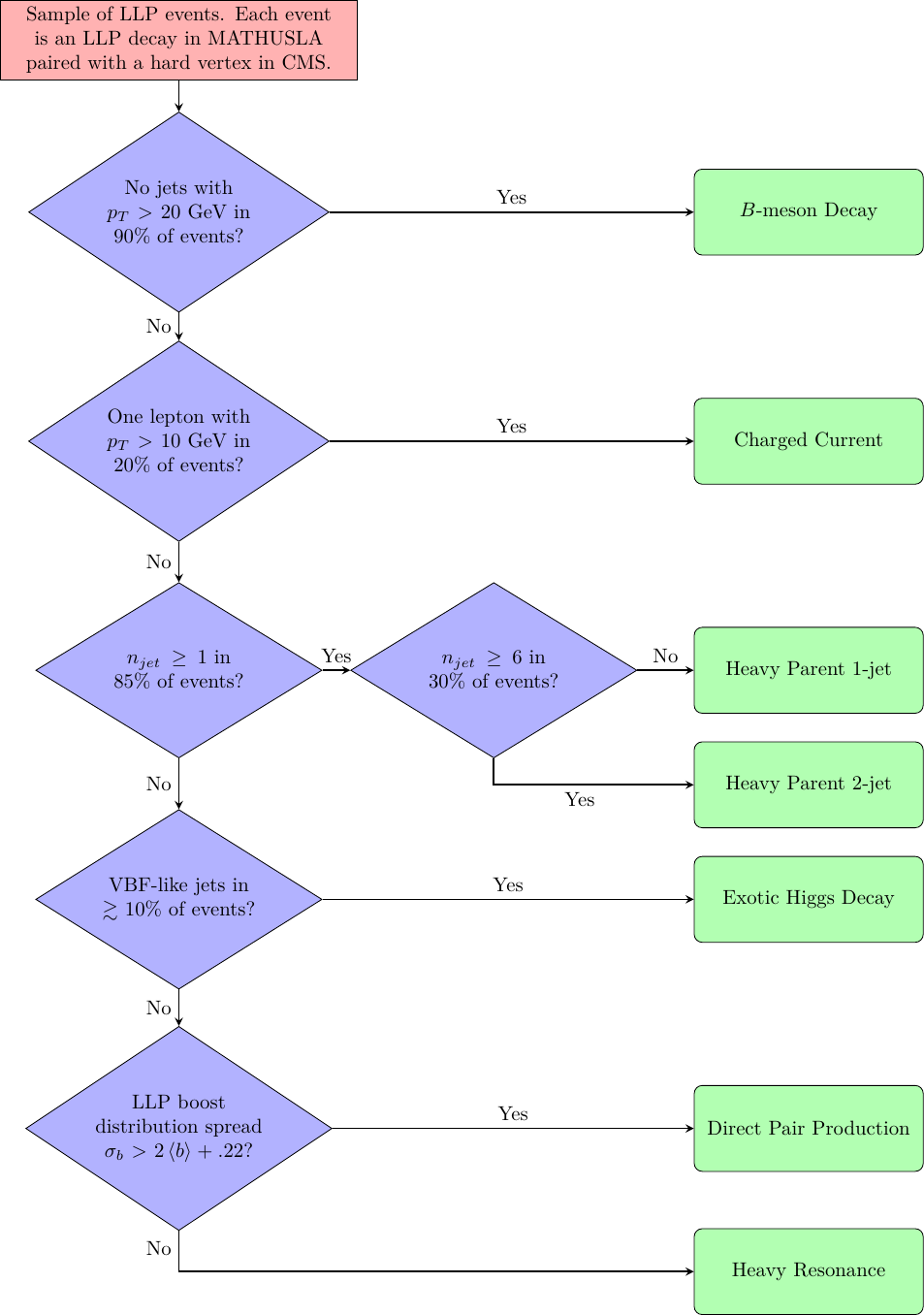}
		\caption{Summary of our hierarchical LLP production mode classification algorithm.}
		\label{fig:Flowchart}
	\end{figure}

	Our goal is to find robust, physically motivated observables that allow us to classify samples of LLP production events into the categories defined by the different simplified models in Figure~\ref{fig:feynmandiagrams}. 
We therefore construct a hierarchical classification algorithm that addresses each production mode hypothesis in order of increasing  difficulty. It is summarized in Figure~\ref{fig:Flowchart} and Table~\ref{tab:classificationalgorithm}. Below, each step is discussed in detail.

		\begin{table}
		\centering
		\hspace*{-6mm}
		\begin{tabular}{|m{5cm}|m{5cm}|c|}
			\hline
			Model & Variable & Cut for  model classification \\
			\hline\hline
			$B$ decay (BB) & Fraction of events with no jets of $p_{T}>20\ \GeV$ & >0.9\\\hline
			Charged Current (CC) & Fraction of events with one lepton of $p_{T}>10\ \GeV$ & >0.2\\\hline
			Heavy Parent (LLP +1 jet)& Fraction of events with at least one jet of $p_{T}>20\ \GeV$ & >0.85 \\\hline
			Heavy Parent (LLP +2 jet)& Fraction of events with at least six jets with $p_{T}>20\ \GeV$ & >0.3\\\hline
			\vspace*{5mm}
			\multirow{2}{*}{Exotic Higgs Decay} & Fraction of events y with highest jet $p_{T}>90\ \GeV$ & 0.04 < y < 0.22\\ \cline{2-3}
			 & Fraction of events x with two jets with $\Delta\eta_{jj}>2.5$ & y < 3.5x - 0.2 \\\hline
			Direct Pair Production & Standard deviation $\sigma$ and mean $\mu$ of LLP $\log_{10}(b)$ distribution & $\sigma>2\mu + .22$ \\\hline
			Heavy Resonance  & --- & ---\\
			\hline
		\end{tabular}
		\caption{Cuts defining the LLP production model classification algorithm, applied in sequence from top to bottom. An observed event sample  that fails all cuts is classified as Heavy Resonance production mode by a process of elimination.}
		\label{tab:classificationalgorithm}	
	\end{table}
	
		The two easiest tasks are to identify data arising from the Exotic $B$-meson Decay and Charged Current models. We start with BB. Because the vast majority of $B$-mesons are produced in relatively soft $b \bar b$ events at the LHC with $p_{T} < 20$ GeV, we can expect very few jets and even fewer b-tags in events from this model. Only $\approx 0.5\%$ of events have a b-tagged jet with $p_{T} > 20$ GeV in the Delphes simulation. The presence of a b-tag could thus only be useful when $\mathcal{O}(1000)$ LLPs have been detected. %
However, there is an upside to the low characteristic energy scale of these events - a complete absence of hard jets in $>90\%$ of events. All of the other models under consideration have a hard process at some high scale, producing ISR jets at the least in a sizeable fraction of events. The cut used to classify a sample as BB-like is essentially a jet veto, demanding zero jets above a certain $p_{T}^{min}$ cutoff in a high fraction of events. 
If the event sample fails this jet veto, we move on to the next possibility.

Next, we check whether the sample could arise from the Charged Current model. In this model, the LLP produced is accompanied by a SM charged lepton. For most values of $m_{LLP}$ and $m_{parent}$, this lepton is generically produced at high transverse momentum, and in the opposite azimuthal direction from MATHUSLA for events where the LLP is detected at MATHUSLA. None of the other simplified models have a mechanism to produce a \emph{single} hard SM lepton in such a high fraction of events. The cut for this model will demand an isolated lepton above some minimum $p_{T}^{min}$ in a high fraction of events. 

The next step in the classification is to check for the Heavy Parent model. In the other models featuring LLPs produced in the decay of a heavy particle, the only jets in an LLP production event come from initial state radiation (with the exception of VBF jets in Higgs production). The spectrum of this radiation is dependent on the characteristic energy scale (and therefore BSM particle masses) of each model, but in general the number of jets per event and their $p_T$ is lower than for the HP model, where we assume the SM products of the heavy parent decay give rise to jets. To be classified as HP, some minimum fraction of events in a sample will have to contain one or more jets above some $p_{T}^{min}$. Within the HP model, we can further distinguish between the decay of the heavy parent particle to one SM particle plus an LLP, and the decay to two SM particles plus an LLP. This secondary classification asks for an even higher jet multiplicity in some minimum fraction of events, and may take advantage of the fact that 4 hard quarks or gluons at parton-level may be quite likely to split into even more jets after showering.

\begin{figure}
\begin{center}
\begin{tabular}{cc}
\includegraphics[width=.45 \textwidth]{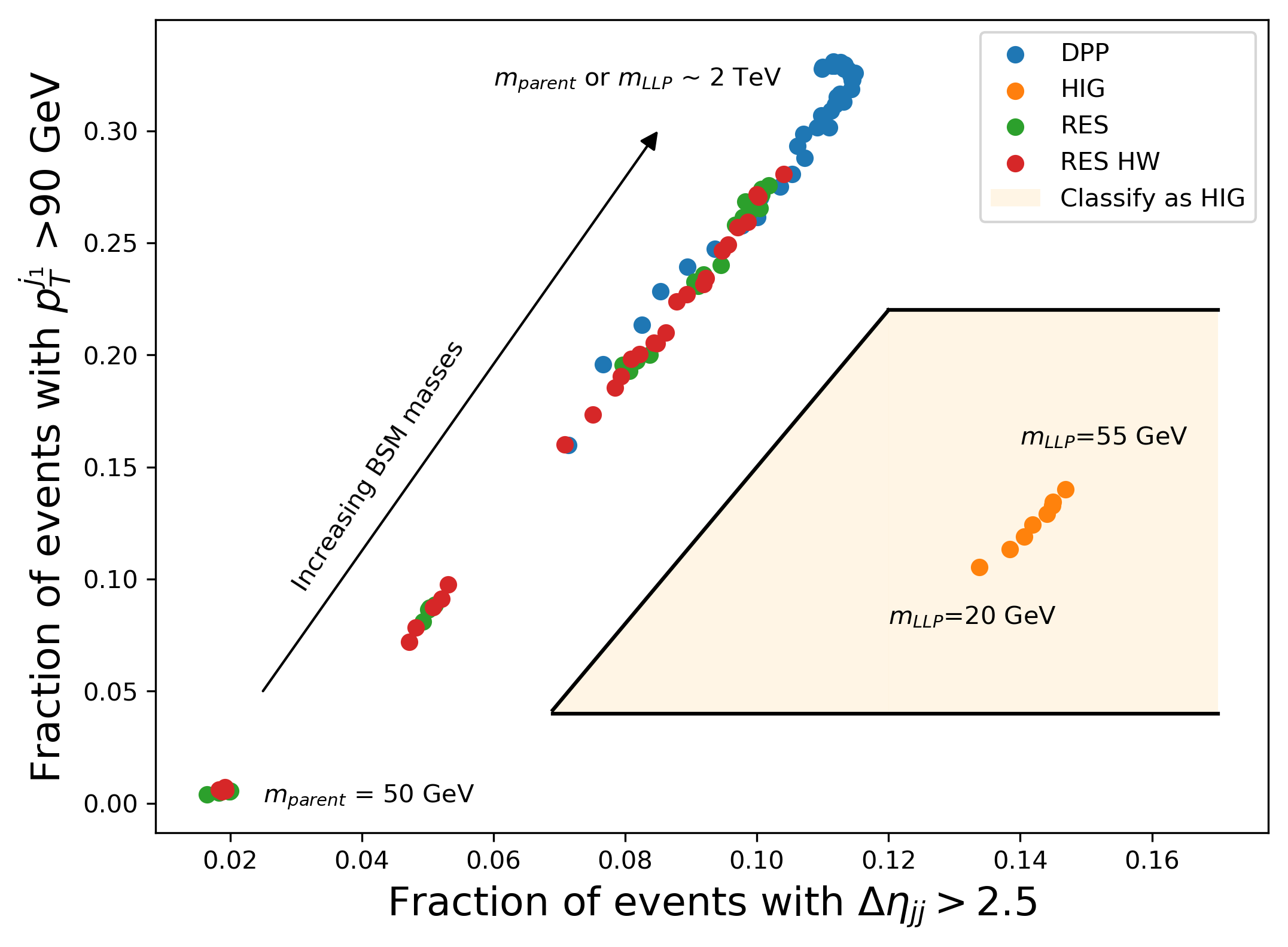}
& 
\includegraphics[width=.45 \textwidth]{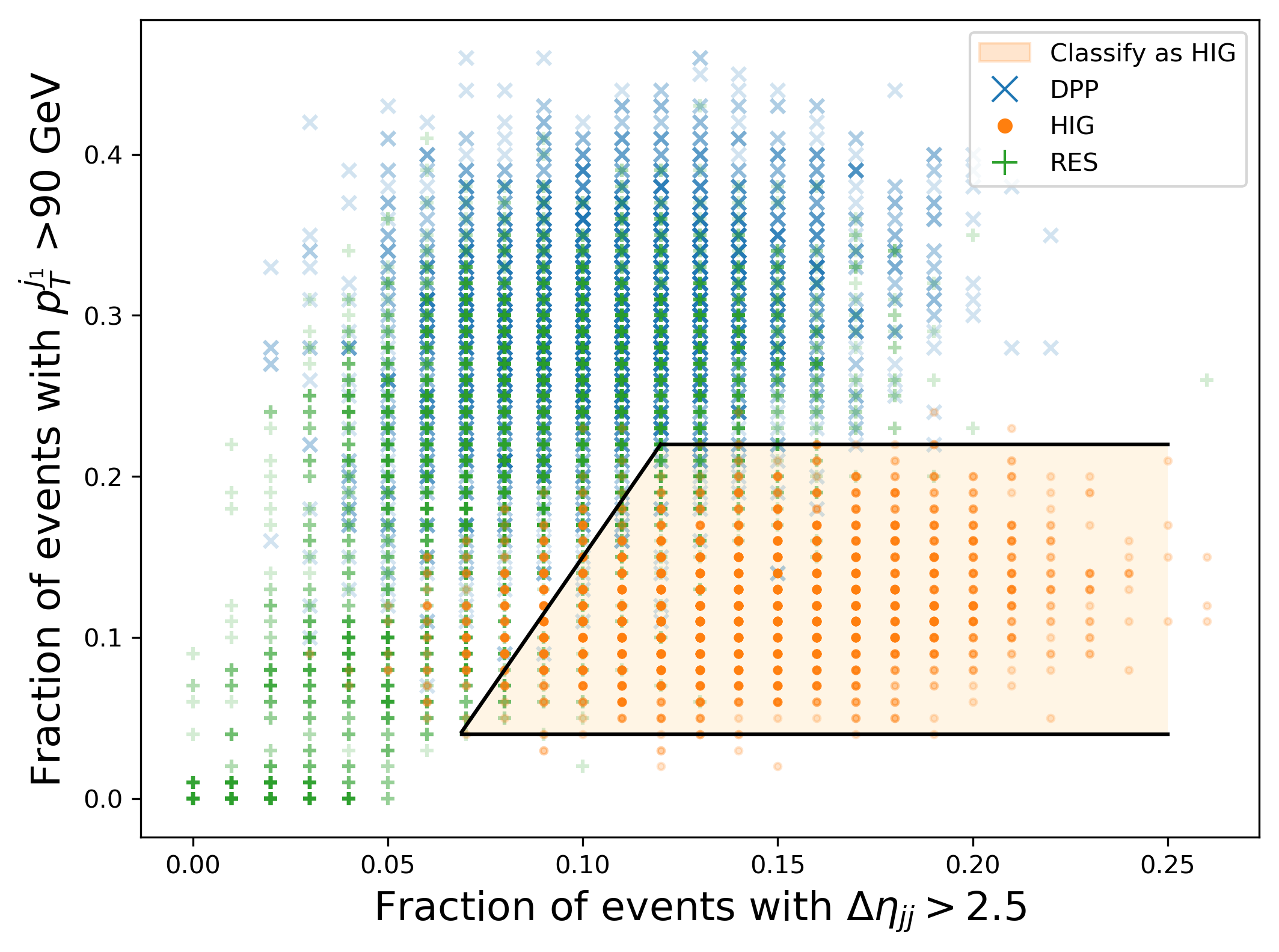}
\end{tabular}
\end{center}
\vspace*{-5mm}
\caption{For Higgs Decay, Heavy Resonance and Direct Pair Production simulation samples, for the range of masses in Table~\ref{tab:simparams}, the left plot shows for each sample the fraction of events with $\Delta n_{jj} > 2.5$ vs the fraction of events with hardest jet $p_T$ above 90 GeV. Clearly, there is very good separation between HIG and RES/DPP in the high-statistics limit. The right plot shows the distribution of these two variables for samples of size $N_{obs} = 100$ for the various production modes and masses. There is now significantly more spread, but HIG events are still well separated. The orange shaded region defines our cuts the sample must pass to be classified as Higgs production.}
\label{fig:HIGcuts}
\end{figure}

After these easier classification decisions, we must turn to more difficult cases requiring slightly more sophisticated checks. The Heavy Resonance and Direct Pair Production modes produce no associated SM particles in the hard process that creates the LLP, and the same is true in the dominant gluon fusion channel for Higgs production. However, in the vector boson fusion (VBF) and associated vector boson (VH) production channels for Higgs production, there are additional particles created in the hard process. In particular, we will look for the signature of VBF events to distinguish the HIG model. 
These events are characterized by a pair of hard jets well-separated in pseudorapidity ($\Delta \eta_{jj}$), originating from the two quarks surviving from the VBF process. Since VBF accounts for approximately 10\% of the Higgs production cross-section, we expect to see roughly the same fraction of Higgs events with this distinctive signature. The DPP and RES models are much less likely to produce events of this nature. Apart from this, the jets present in HIG events have a weaker $p_{T}$ spectrum than for RES or DPP in most regions of parameter space, due to the heavy resonance having mass $> m_{H}$ or the produced LLPs having $2m_{LLP} > m_{H}$, controlling the scale of the interaction and therefore the momentum of produced ISR jets. With both of these considerations in mind, we can construct a combined cut in the fraction of events with highest jet $p_{T}$ larger than some cutoff, along with fraction of events with $\Delta \eta _{jj}$ greater than some cutoff. See Figure~\ref{fig:HIGcuts} for the shape of this cut, along with the high-statistics prediction of these two variables for the HIG, RES and DPP samples with various masses, and the spread for observed samples of size $N_{obs} = 100$. 

 Finally, we must devise a strategy to distinguish between the two most similar models, Heavy Resonance and Direct Pair Production. Indeed, in the high $m_{parent}$ limit, the RES and DPP models coincide. For the first time, we will use the information gathered by MATHUSLA for production mode classification, not just the fact that it triggers CMS. In the RES model, the boost of the detected LLPs is determined almost exclusively by $m_{LLP}/m_{parent}$, since the heavy resonant particles will be produced dominantly on-threshold, giving the LLPs a boost $b \sim m_{parent}/2 m_{LLP}$. This produces a fairly narrow boost distribution for the LLPs if the heavy resonance has a narrow width. Contrastingly, in the DPP model the LLP boost is correlated directly with $m_{LLP}$, with higher masses corresponding to lower boosts. Because there is no resonance decaying to the LLPs in the DPP model, the energy distribution of the LLPs is wider, leading to a wider boost distribution compared to even a high-width heavy resonance. 
		
\begin{figure}
	\begin{center}
	\begin{tabular}{cc}
	\includegraphics[width=.45\textwidth]{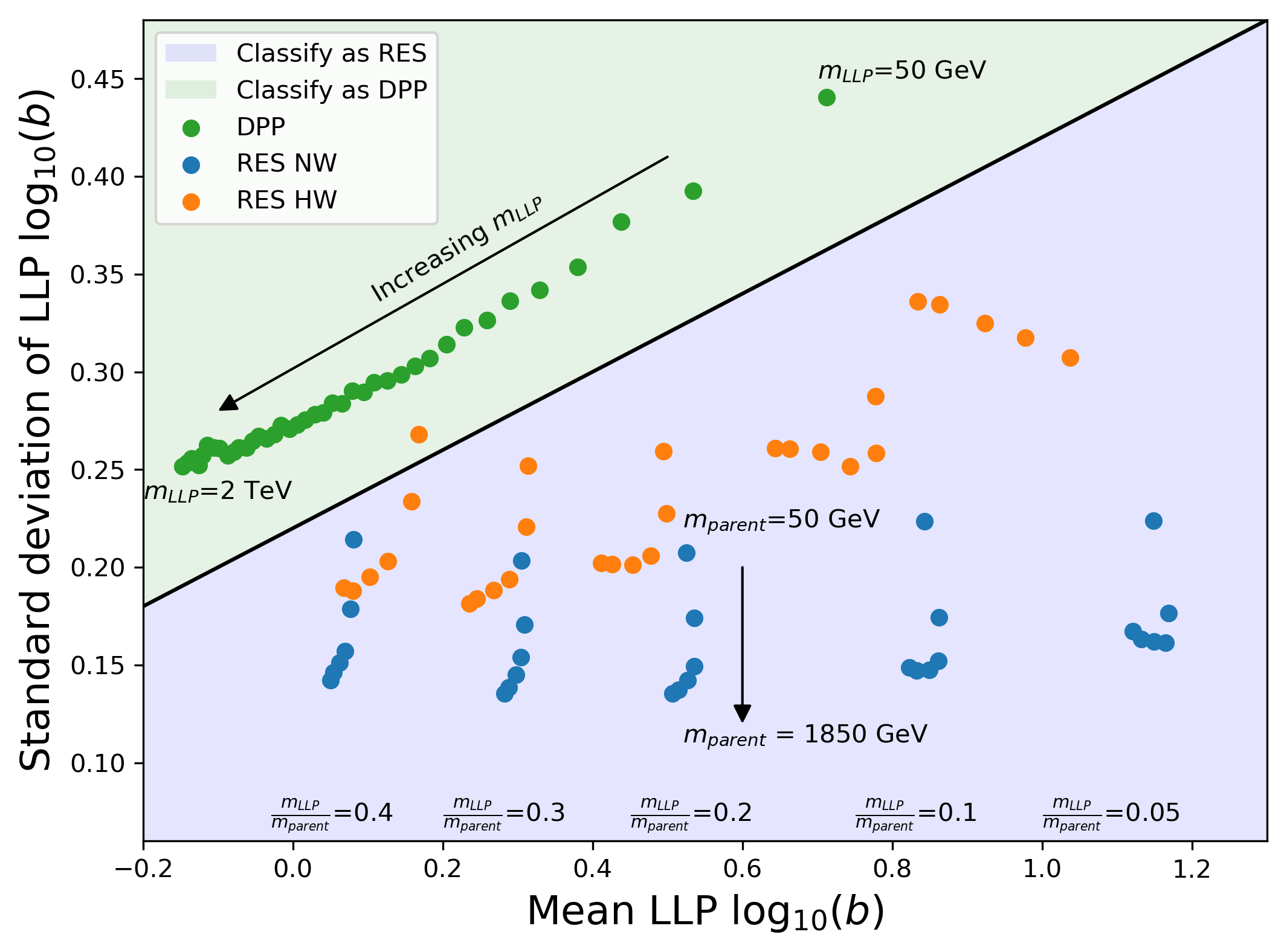}
	&
	\includegraphics[width=.45\textwidth]{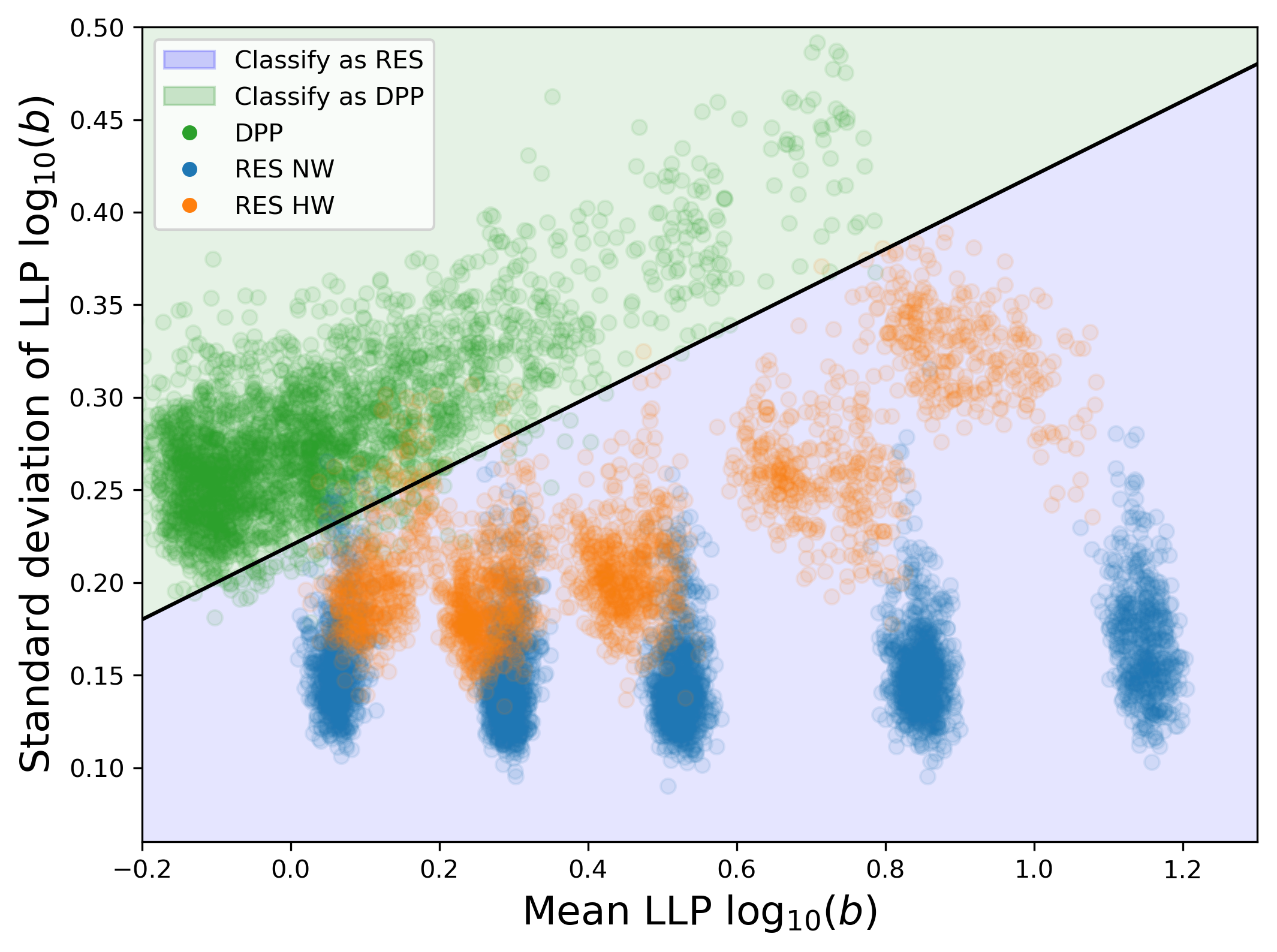}
	\end{tabular}
	\end{center}
	\vspace*{-5mm}
	\caption{LLP boost mean and standard deviation of the low- and high-width Heavy Resonance samples for the various simulated masses, compared to the Direct Pair Production samples. Left: high-statistics limit. Right: for observed samples of size $N_{obs} = 100$. The indicated diagonal cut in this plane can separate RES from DPP samples. We also indicate the parent and LLP masses of each high statistics sample. Gaps in the RES distribution are due to the coarse grid of simulated masses.}
	\label{fig:DPPCuts}
	\end{figure}

The difference in correlations of the LLP boost with particle masses between the two models will allow us to define a cut that separates the two models. Figure~\ref{fig:DPPCuts} shows the LLP boost mean and standard deviation of the low- and high-width Heavy Resonance samples for the various simulated masses, compared to the Direct Pair Production samples, in both the high-statistics limit and for observed samples of size $N_{obs} = 100$. 
A diagonal cut in this plane can separate DPP and RES.  This cut only loses some of its distinguishing power in the high-width limit for the RES model. \footnote{For the initial classification we will make no effort to distinguish between the low- and high-width limits of the RES model. Since the width of the heavy resonant particle is a continuous parameter, the task of determining the width is better framed as part of parameter estimation. Section \ref{subsection:resparamestimation} sketches how this could be done.}

For all of the above cuts we optimize the precise thresholds to maximize classification performance for event samples of size $N_{obs} = 100$. The resulting cuts are summarized in Table \ref{tab:classificationalgorithm}, and visualized as a flowchart in Figure~\ref{fig:Flowchart}. Note that these cuts are applied in order, such that only samples failing to meet the criteria of a given step are tested in the next. 
While more sophisticated simulations or data-driven optimizations might result in slightly different optimal values of these cuts, their physical basis nevertheless makes them robust.

\begin{figure}
	\centering
	\includegraphics[width=.9\textwidth,scale=0.95]{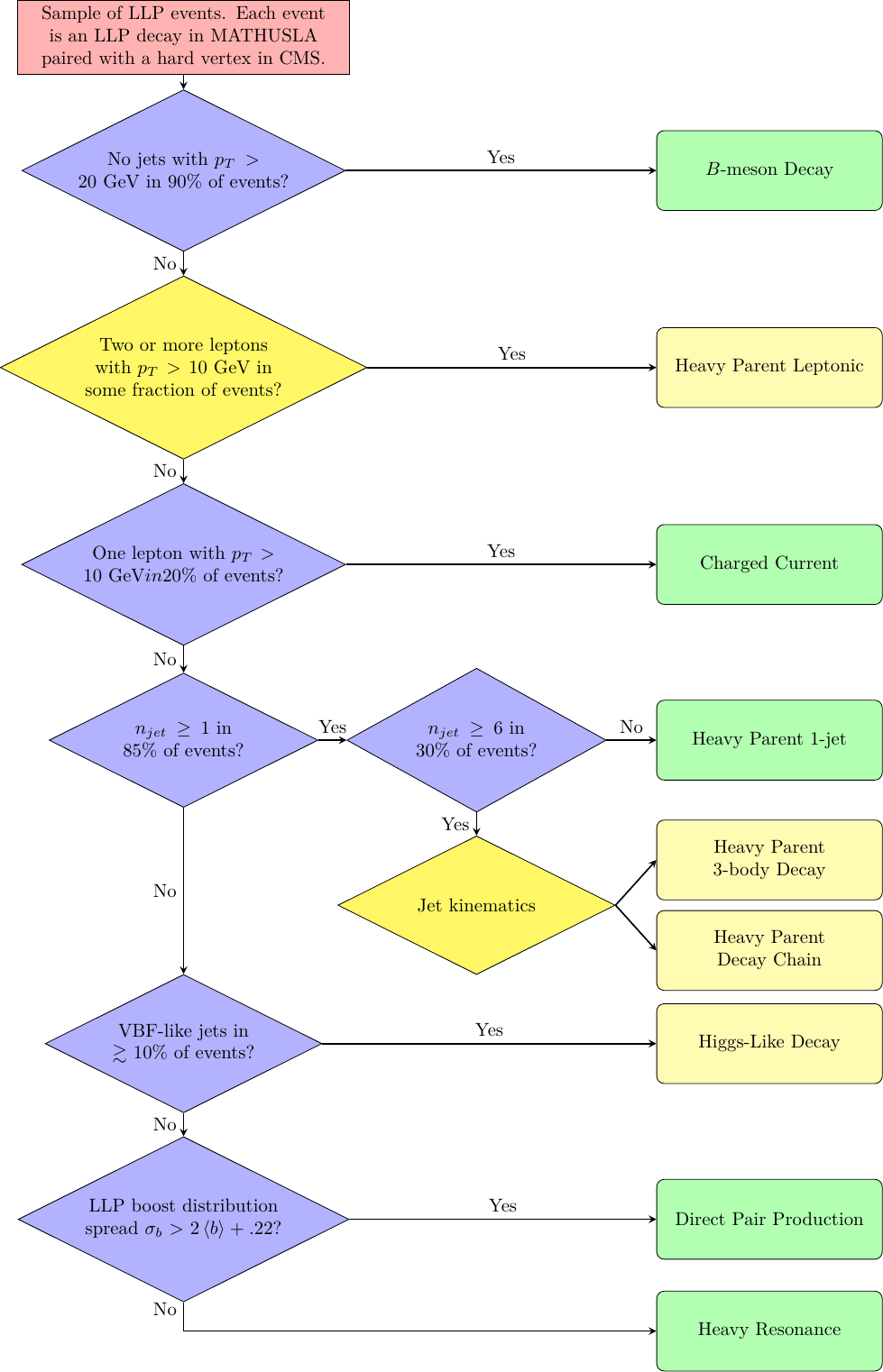}
	\caption{Flow chart summary of our hierarchical LLP production mode classification algorithm. showing possible additions to include other LLP production scenarios not considered in our detailed analysis.}
	\label{fig:Flowchart2}
\end{figure}
	
Our classification framework is easily extendable to include other LLP scenarios. 
For example, leptonic decay in the Heavy Parent model can be accommodated by testing for a minimum fraction of events with two or more hard leptons. 
A modification of the two-jet Heavy Parent scenario where the heavy parent undergoes a decay chain rather than three-body decay could be distinguished using jet kinematic variables like $M_{T2}$~\cite{Burns:2008va}. 
Finally, a generalized Higgs decay scenario with indeterminate mass for the Higgs-like scalar could still be identified by the presence of VBF production events,\footnote{A singlet scalar without electroweak gauge charge would be difficult to distinguish from the $Z'$-like Heavy Resonance model if only LLP production events are considered, but if such a state existed, the pattern of its visible decays in the main detector could distinguish gauge-ordered from Yukawa-ordered couplings and hence favor a vector or scalar interpretation.} with the task of distinguishing the scalar resonance from the SM Higgs becoming a problem of parameter estimation similar to the cases we discuss in Section~\ref{section:paramestimation}.
We have omitted these scenarios from our first pilot analysis for simplicity, or because they are easier to classify than the cases we study (in particular the leptonic heavy-parent models). Even so, we can anticipate how our classification algorithm would be modified to include these possibilities in Figure~\ref{fig:Flowchart2}. We hope this will be useful for future extensions of our analysis. 

It is important to note that nothing about our methods requires the LLP to be detected in an external detector like MATHUSLA. If LLPs are discovered at the LHC main detectors, the same kind of analysis can be used to characterize them as well. In fact, more information will be available, since the main detectors might be able to extract energy and momentum data on the LLP decay products, depending on the position of the DV within the detector.

\subsection{Performance}

Unweighted sets of pseudo-data with $N_{obs}=10,100,1000$ were generated 
for the models and mass ranges shown in Table~\ref{tab:simparams}. For most models, 30-40 parameter-points were generated, forming a rough grid in the $m_{parent},m_{LLP}/m_{parent}$ plane. For the one-parameter models, between 8 (HIG) and 40 (DPP) values of $m_{LLP}$ were generated.
	
The classification algorithm was run on all of these pseudo-data samples, and the rates of correct and incorrect classifications were recorded. At each of the three benchmark values $N_{obs}$, we report an average classification accuracy for each model. This average of classification accuracies is taken first over all samples at a given point in parameter space, then over all points in parameter space, for each model. This average is particular to the prior imposed by our choice of simulated BSM particle masses, but generally reflects the ability of the classifier to distinguish a given model from the others across parameter space. Corners of parameter space for which the classification accuracy differs significantly from the average are discussed on a case-by-case basis, as are trends of variation in the accuracy as a function of BSM particle masses and widths.

	\begin{table}
		\centering
		\begin{tabular}{|c|c|c|c|}
			\hline
			$N_{obs}$ & 10 & 100 & 1000 \\
			\hline
			$B$-decay (BB) & 98&100 & 100 \\
			Charged Current (CC) & 94& 98 & 98\\
			Heavy Parent (HP) & 91& 92& 93\\
			Higgs Decay (HIG) &36 & 91& 100 \\
			Direct Pair Production (DPP) & 48& 98& 100\\
			Resonance, narrow-width (RES NW)&71 &93 &98 \\
			Resonance, high-width (RES HW) & 63&86 & 94 \\
			\hline
		\end{tabular}
		\caption{Classification accuracies (\%) for each model at varying $N_{obs}$, averaged over all simulated LLP and parent particle masses for each model. For more details, see Table~\ref{tab:classificationtable}.}
		\label{tab:classificationresults}	
	\end{table}

	\begin{table}
		\centering
		
		\hspace*{-9mm}
		\begin{tabular}{l}
		
		$\mathbf{N_{obs} = 10}$:
		\\
		\begin{tabular}
		{|m{3cm}|m{1.4cm}|m{1.4cm}|m{1.4cm}|m{1.4cm}|m{1.4cm}|m{1.4cm}|m{1.4cm}|}
			\hline
			\backslashbox{Truth}{Result} & BB & CC & HP 1-jet& HP 2-jet & HIG & DPP & RES \\
			\hline
			BB (4570) &$ 98.4^{+0.3}_{-0.4}$ &0 &0 &0 &0 &$ 1.4^{+0.4}_{-0.3}$ &$ 0.2^{+0.2}_{-0.1}$\\CC (20920) &$ 4.0^{+0.3}_{-0.3}$ &$ 93.5^{+0.3}_{-0.3}$ &$ 0.004^{+.017}_{-.004}$ &0 &$ 0.03^{+0.03}_{-0.02}$ &$ 2.2^{+0.2}_{-0.2}$ &$ 0.2^{+0.1}_{-0.1}$\\HP 1-jet (26510) &$ 0.1^{+0.04}_{-0.03}$ &0 &$ 81.8^{+0.5}_{-0.5}$ &$ 9.2^{+0.4}_{-0.3}$ &$ 1.3^{+0.1}_{-0.1}$ &$ 5.7^{+0.3}_{-0.3}$ &$ 2.0^{+0.2}_{-0.2}$\\HP 2-jet (12360) &0 &0 &$ 19.9^{+0.7}_{-0.7}$ &$ 78.8^{+0.7}_{-0.7}$ &$ 0.5^{+0.1}_{-0.1}$ &$ 0.8^{+0.2}_{-0.1}$ &$ 0.1^{+0.1}_{-0.0}$\\HIG (28550) &$ 0.02^{+0.02}_{-0.01}$ &$ 0.04^{+0.03}_{-0.02}$ &$ 9.5^{+0.3}_{-0.3}$ &0 &$ 36.1^{+0.6}_{-0.6}$ &$ 29.4^{+0.5}_{-0.5}$ &$ 24.9^{+0.5}_{-0.5}$\\DPP (29120) &$ 0.01^{+0.02}_{-0.01}$ &0 &$ 17.5^{+0.4}_{-0.4}$ &0 &$ 12.3^{+0.4}_{-0.4}$ &$ 48.0^{+0.6}_{-0.6}$ &$ 22.2^{+0.5}_{-0.5}$\\RES NW (40160) &$ 4.7^{+0.2}_{-0.2}$ &0 &$ 7.6^{+0.3}_{-0.3}$ &0 &$ 14.0^{+0.3}_{-0.3}$ &$ 3.0^{+0.2}_{-0.2}$ &$ 70.7^{+0.4}_{-0.4}$\\RES HW (23100) &$ 3.9^{+0.3}_{-0.2}$ &0 &$ 6.6^{+0.3}_{-0.3}$ &0 &$ 14.9^{+0.5}_{-0.5}$ &$ 11.4^{+0.4}_{-0.4}$ &$ 63.1^{+0.6}_{-0.6}$\\
			\hline
		\end{tabular}
		\\ \\
		$\mathbf{N_{obs} = 100}$:
		\\		
		\begin{tabular}{|m{3cm}|m{1.4cm}|m{1.4cm}|m{1.4cm}|m{1.4cm}|m{1.4cm}|m{1.4cm}|m{1.4cm}|}
			\hline
			\backslashbox{Truth}{Result}& BB & CC & HP 1-jet& HP 2-jet & HIG & DPP & RES \\
			\hline
			BB (457) &$ 100.0^{+0.0}_{-0.5}$ &0 &0 &0 &0 &0 &0\\CC (2092) &0 &$ 97.6^{+0.6}_{-0.7}$ &0 &0 &0 &$ 2.4^{+0.7}_{-0.6}$ &0\\HP 1-jet (2651) &0 &0 &$ 90.5^{+1.1}_{-1.2}$ &$ 0.9^{+0.4}_{-0.3}$ &$ 1.0^{+0.4}_{-0.3}$ &$ 7.3^{+1.0}_{-0.9}$ &$ 0.3^{+0.3}_{-0.2}$\\HP 2-jet (1236) &0 &0 &$ 16.0^{+2.1}_{-2.0}$ &$ 83.5^{+2.0}_{-2.1}$ &$ 0.4^{+0.5}_{-0.2}$ &$ 0.1^{+0.3}_{-0.1}$ &0\\HIG (2855) &0 &0 &0 &0 &$ 90.4^{+1.0}_{-1.1}$ &$ 7.4^{+1.0}_{-0.9}$ &$ 2.2^{+0.6}_{-0.5}$\\DPP (2912) &0 &0 &$ 0.1^{+0.2}_{-0.1}$ &0 &$ 1.6^{+0.5}_{-0.4}$ &$ 97.9^{+0.5}_{-0.6}$ &$ 0.4^{+0.3}_{-0.2}$\\RES NW (4016) &0 &0 &0 &0 &$ 4.8^{+0.7}_{-0.6}$ &$ 1.9^{+0.5}_{-0.4}$ &$ 93.3^{+0.7}_{-0.8}$\\RES HW (2310) &0 &0 &0 &0 &$ 6.3^{+1.0}_{-0.9}$ &$ 8.0^{+1.2}_{-1.1}$ &$ 85.7^{+1.4}_{-1.5}$\\
			\hline
		\end{tabular}
		
		\\ \\
		$\mathbf{N_{obs} = 1000}$:
		\\		

		\begin{tabular}{|m{3cm}|m{1.4cm}|m{1.4cm}|m{1.4cm}|m{1.4cm}|m{1.4cm}|m{1.4cm}|m{1.4cm}|}
			\hline
			\backslashbox{Truth}{Result} & BB & CC & HP 1-jet& HP 2-jet & HIG & DPP & RES\\
			\hline
			BB (40) &$ 100.0^{+0.0}_{-6.0}$ &0 &0 &0 &0 &0 &0\\CC (191) &0 &$ 97.6^{+1.5}_{-3.0}$ &0 &0 &0 &$ 2.4^{+3.0}_{-1.5}$ &0\\HP 1-jet (246) &0 &0 &$ 92.7^{+2.8}_{-3.8}$ &0 &0 &$ 7.3^{+3.8}_{-2.8}$ &0\\HP 2-jet (112) &0 &0 &$ 15.0^{+7.5}_{-5.7}$ &$ 85.0^{+5.7}_{-7.5}$ &0 &0 &0\\HIG (282) &0 &0 &0 &0 &$ 100.0^{+0.0}_{-0.9}$ &0 &0\\DPP (274) &0 &0 &0 &0 &0 &$ 100.0^{+0.0}_{-0.9}$ &0\\RES NW (390) &0 &0 &0 &0 &0 &$ 1.7^{+1.7}_{-0.9}$ &$ 98.3^{+0.9}_{-1.7}$\\RES HW (219) &0 &0 &0 &0 &$ 0.3^{+1.5}_{-0.2}$ &$ 5.7^{+3.7}_{-2.5}$ &$ 94.0^{+2.6}_{-3.7}$\\
			\hline
		\end{tabular}
		
		\end{tabular}
	
		\caption{Breakdown of the Production Mode Classifier output, for pseudo-data samples with 10, 100 or 1000 events, averaged over all LLP and parent particle masses simulated for each model. Entries in row i, column j show the percentage of samples from model i classified as model j. 95\% confidence intervals are shown for non-zero classification accuracies, accounting only for statistical uncertainty due to the limited number of samples tested. The number of samples tested for each model is listed in brackets in the first column.} 
		\label{tab:classificationtable}	
	\end{table} 
	
	Table \ref{tab:classificationresults} summarizes the performance of the classifier for each model, while Table~\ref{tab:classificationtable} shows the full classification matrices for each benchmark $N_{obs}$. 
	\begin{enumerate}
	\item All pseudo-experiments of BB data with $N_{obs}=100$ and 1000 were correctly classified, and no pseudo-experiments were incorrectly classified as BB. There were 29 samples of 100 events per $m_{LLP}$ point on average, and 3 samples of 1000 events on average. The limited statistics available is not concerning, since the classification accuracy reaches 100\% already at $N_{obs}=100$. 
	
	\item	In the CC model, all pseudo-experiments with $N_{obs}=100$ were correctly classified except for those with $m_{parent}=50\ \GeV$ and $m_{LLP}=45\ \GeV$. No pseudo-experiments from other models were incorrectly classified as CC. That performance persisted for $N_{obs}=1000$. For $N_{obs}=10$, performance degraded for $250\ \GeV\leq m_{parent}<1\ \TeV$ down to $\approx 97-99\%$ classification accuracy, and  for $m_{parent}=50\ \GeV$, classification accuracies varied between 0-70\% for $m_{LLP}$ ranging from 45 GeV down to 5 GeV. Still, no samples from other models were incorrectly classified as CC. 
	
	\item   For $m_{parent}\geq 250\ \GeV$ in the HP model, all pseudo-experiments with $N_{obs}\geq 100$ were classified as HP, whether in the one or two jet decay modes. There were $\mathcal{O}(50)$ pseudo-data samples tested per point in parameter space. For $m_{parent}=50\ \GeV$, the lowest simulated parent particle mass, classification accuracy ranged from 88\% at $m_{LLP}/m_{parent}=0.1$ down to 0\% for $m_{LLP}/m_{parent} > 0.5$ . With such low available energy for SM products in these events, this low classification accuracy reflects the physical similarity of the experimental signature of the HP model to the signatures of the other simplified models in the low-mass corner of parameter space. This pattern of performance persisted from $N_{obs}=10$ to $N_{obs}=1000$. Above $m_{parent}\approx 500\ \GeV$, the one- and two-jet decay modes were completely distinguishable. At $m_{parent}=200\ \GeV$, a significant fraction of HP 2-jet samples were classified as HP 1-jet, ranging from 100\% incorrectly classified at $m_{LLP}=180\ \GeV$ down to 18\% at $m_{LLP}=20\ \GeV$.

	\item	For the HIG model, 91\% of pseudo-experiments of $N_{obs}=100$ events were correctly classified. This performance is indepenent of $m_{LLP}$. 
	
	\item	For the DPP model, correct classification rates of pseudo-experiments with $N_{obs}=100$ varied from $90\%$ at $m_{LLP}=50\ \GeV$ up to approximately $\approx 98-100\%$ for $m_{LLP} \geq 700\ \GeV$. The number of unweighted samples of 100 events available varied from 20 at low $m_{LLP} \sim 50\ \GeV$ to 80 at high $m_{LLP}\gtrsim 500\ \GeV$, with a total of 2275 samples tested. For both the HP and CC models with $m_{parent}\sim 50\ \GeV,m_{LLP}\sim 45\ \GeV$, most samples were incorrectly classified as DPP. 
	
	\item	For the RES model with a narrow-width resonance, the average classification accuracy of 100 event pseudo-data samples across parameter space was 93\%. Accuracies varied from $\approx 85-100\%$ across parameter space, with the exception of $m_{parent} = 50\ \GeV$, $m_{LLP}/m_{parent}=0.4$, where 52\% of samples were mis-classified as DPP. Generally, classification accuracy rose with $m_{parent}$. 
	
	\item	For the RES model with a high-width resonance ($\Gamma_{Z'} = 0.3\times m_{Z'}$), classification accuracy was lower than for narrow-width, averaging $86\%$ across the simulated parameter space. The variation of classification accuracy across parameter space was similar to the narrow-width case. 
	\end{enumerate}

Clearly, our classification algorithm performs well in most regions of BSM parameter space. A small minority of scenarios is misclassified, but in each such case this reflects either a genuine physical degeneracy of observables, like in various low-mass limits, requiring much more detailed analysis, or simply a `special case', where an extension of our simple classifier could greatly improve accuracy. Having demonstrated that classification is generally robust, we leave such further details to future studies.

\section{Determining Model Parameters}
	\label{section:paramestimation}
	
The second task for which we would like to estimate the prospective capabilities of MATHUSLA and CMS is the measurement of the properties of the newly discovered BSM particles. In Ref.~\cite{Curtin:2017izq} it was shown that the measured LLP boost distribution gives information about the LLP mass under the assumption of production by exotic Higgs decay. We will attempt to demonstrate similar relations for the rest of the simplified models under our consideration. Without the knowledge of an intermediate parent particle whose mass is itself known, this task becomes more difficult. Each subsection below will describe the estimators of BSM particle masses and precision achievable for each of the simplified models. 
One important assumption that we make here is that the production mode of the LLPs in the sample has already been accurately identified, so that we know which analysis to perform.\footnote{This assumption could be loosened if we extended our methods to some kind of global template-fit over all BSM models, but this would have to contend with detailed mis-modeling effects which our simple cut-based classifier algorithm mostly avoids.}
The main conclusion of our analysis is that MATHUSLA can determine BSM particle masses with useful $\mathcal{O}(10\%)$ precision assuming only $\mathcal{O}(100)$ observed events. 
	
For each production mode, we use binned maximum-likelihood estimation to compare distributions of event-level observables from pseudo-data with $N_{obs} = 10, 100$ or 1000 events against distributions generated using our full high-statistics simulations. Best-fit masses are computed for many pseudo-data samples, and the standard deviation of the distribution of best-fit masses is taken as an estimate of the precision MATHUSLA can achieve when measuring LLP model parameters. The log-likelihood (LLH) function to be minimized as a function of the theory parameters is 
\begin{equation}
-\ln L(m_{parent}, m_{LLP}) = \sum_{i} n_{i}\ln\left(\frac{n_{i}}{\mu_{i}(m_{parent},m_{LLP})}\right)\
\end{equation}
where $i$ indexes bins in a histogram of the chosen observables, $n_{i}$ is the observed count in bin $i$ for a given sample of pseudo-data, and $\mu_{i}(m_{parent},m_{LLP})$ is the expected count in bin $i$. Bins with 0 observed count are ignored. The likelihood function is evaluated over a grid of points in $(m_{parent},m_{LLP}/m_{parent})$ with available simulation, then interpolated and minimized. For models without varying parent particle masses, simulation is generated for $m_{LLP}$ at regular intervals, and the likelihood depends only on $m_{LLP}$.

\begin{table}
	\centering
	\begin{tabular}{|m{4.5cm}|m{2cm}|m{0.9cm}|m{1cm}|m{2.7cm}|m{1.5cm}|}
		\hline
		Model & $x_{1}$ & $x_{2}$ & $N_{obs}$ & $\frac{m_{LLP}}{m_{parent}}$ or $m_{LLP}$ precision& $m_{parent}$ precision\\
		\hline \hline
		\multirow{3}{*}{B decay} & 	
		\multirow{21}{*}{$\log_{10}(b_{LLP})$} & 
			\multirow{3}{*}{-} & 10& $0.3 - 0.7$ &  \\
							&	&& 100& $0.1 - 0.2$& - \\
							&	&&1000 & $\lesssim 0.05$ &\\
							\cline{1-1} \cline{3-6}
		\multirow{3}{*}{Charged Current} & 	 & 	\multirow{3}{*}{$p_{T}^{\ell}$}&10 & $0.1$ &$0.1$\\
								&&&100 & $0.05$ & $0.02$ \\
							&&	& 1000& $0.01$& $0.01$ \\
							\cline{1-1} \cline{3-6}
		\multirow{3}{*}{Heavy Parent} & & 	\multirow{3}{*}{$H_{T}$} &10 & $0.2$ &$0.2$\\
								&&	&100 &$0.05$ &$0.05$  \\
								&&	&1000 & $0.01$& $0.01$ \\
								\cline{1-1} \cline{3-6}
		\multirow{3}{*}{Exotic Higgs decay} & 	& 	\multirow{3}{*}{-} & 10&$0.15$&  \\
										&&	& 100& $0.05$&-  \\
									&&	&1000 & $0.01$&  \\
									\cline{1-1} \cline{3-6}
		 &	& 	\multirow{3}{*}{-} & 10&$0.3 - 0.5$ &\\
									Direct Pair Production&&	&100 &$0.1-0.2$ &- \\
									&&	&1000 & $.03-.07$& \\
									\cline{1-1} \cline{3-6}
		 &	 & 	\multirow{3}{*}{$n_{jet}$} &10 &$0.07$&\\ 
									Heavy Resonance (narrow)&&	& 100&$0.02$ & \\
									&&	&1000 & $0.01$&$0.15$ \\
									\cline{1-1} \cline{3-6}
		 &	 & 	\multirow{3}{*}{$n_{jet}$} &10 &$0.12$&\\ 
		Heavy Resonance (wide)&&	& 100&$0.05$ & \\
		&&	&1000 & $0.02$&$0.15$ \\
		\hline
	\end{tabular}
	\caption{Summary of parameter estimation performance for all of the simplified models we consider. The variables $x_1, x_2$ chosen to estimate BSM particle masses are listed. The precisions shown are the characteristic standard deviation/mean of best-fit masses for benchmark BSM particle masses that are approximately representative for each model.} 
	\label{tab:paramestsummary}	
\end{table}

	The spacing and size of the grid were chosen to accurately demonstrate the precision achievable at $N_{obs}=100$. This means that for $N_{obs}=10$, some outlier pseudo-experiments do not have best-fit mass values within the grid. In such cases, the reported precision should be regarded as an approximate lower bound on the uncertainty in BSM particle masses MATHUSLA can achieve. For $N_{obs}=1000$ the grid spacing is much larger in some cases than the variation in best-fit masses, meaning that the actual distribution of best-fit masses is not accurately portrayed. However, systematic uncertainties driven by the final experimental design are likely to dominate in this regime, so the exact values obtained for 1000 observed events are less important than the qualitative lesson that the BSM parameter measurement might become systematics dominated. 

	For single-parameter models without a resonant LLP parent particle (Direct Pair Production), or where the parent mass is known ($B$-decay, Higgs decay), we only need a single variable $x_1(m_{LLP}) = b =  |\vec p_{LLP}|/m_{LLP}$, the LLP boost that MATHUSLA can measure from the DV geometry. The binned distribution $\mu_i (m_{LLP})$ in this variable is sufficient to measure LLP mass, since $b \sim m_{parent}/m_{LLP}$. 

	For models with unknown parent particle mass (Charged Current, Heavy Parent and Heavy Resonance), we also make use of the LLP boost $x_1(m_{LLP}, m_{parent}) = b$. However, we need to define an additional variable $x_2(m_{LLP}, m_{parent})$ that can be constructed from information at MATHUSLA and CMS, and which supplies information about the LLP or parent particle masses independent of the boost. We do this case-by-case below, and the variables are summarized in Table \ref{tab:paramestsummary}, along with the approximate precision MATHUSLA can achieve for each model. 

	Due to the high computational cost of generating simulated datasets on the finely-spaced grid of LLP and parent particle masses necessary to accurately maximize the likelihood over parameter space, we demonstrate parameter estimation performance for one set of BSM particle masses in the two-parameter models. We choose $m_{parent}=1\ \TeV$, $m_{LLP}=300\ \GeV$ for all models as a benchmark that is fairly representative of MATHUSLA's abilities across a large range of parameter space.

\subsection{Exotic $B$-Meson Decay}
 	The first model we consider is Exotic $B$-meson Decay, in the two cases of a heavy neutral lepton (HNL) LLP or a scalar LLP. We assume these can be distinguished by examining the decay products of observed LLPs. In the HNL case, the LLP decay includes an invisible neutrino, so the visible decay products in MATHUSLA do not point back at the LHC interaction point. The details of how this classification could be accomplished would require careful further study as well as detailed knowledge on the final experimental design of MATHUSLA, which is not currently available. Furthermore, the boost measurement requires special treatment when the LLP mass is not much greater than that of its decay products. These issues are important and require their own dedicated investigation, which is beyond the scope of this work. For our purposes, we will assume that  that the LLP decay mode can be determined and its boost measured with useful accuracy.
 	
\begin{figure}
 		\centering
 		\includegraphics[width=.6\textwidth]{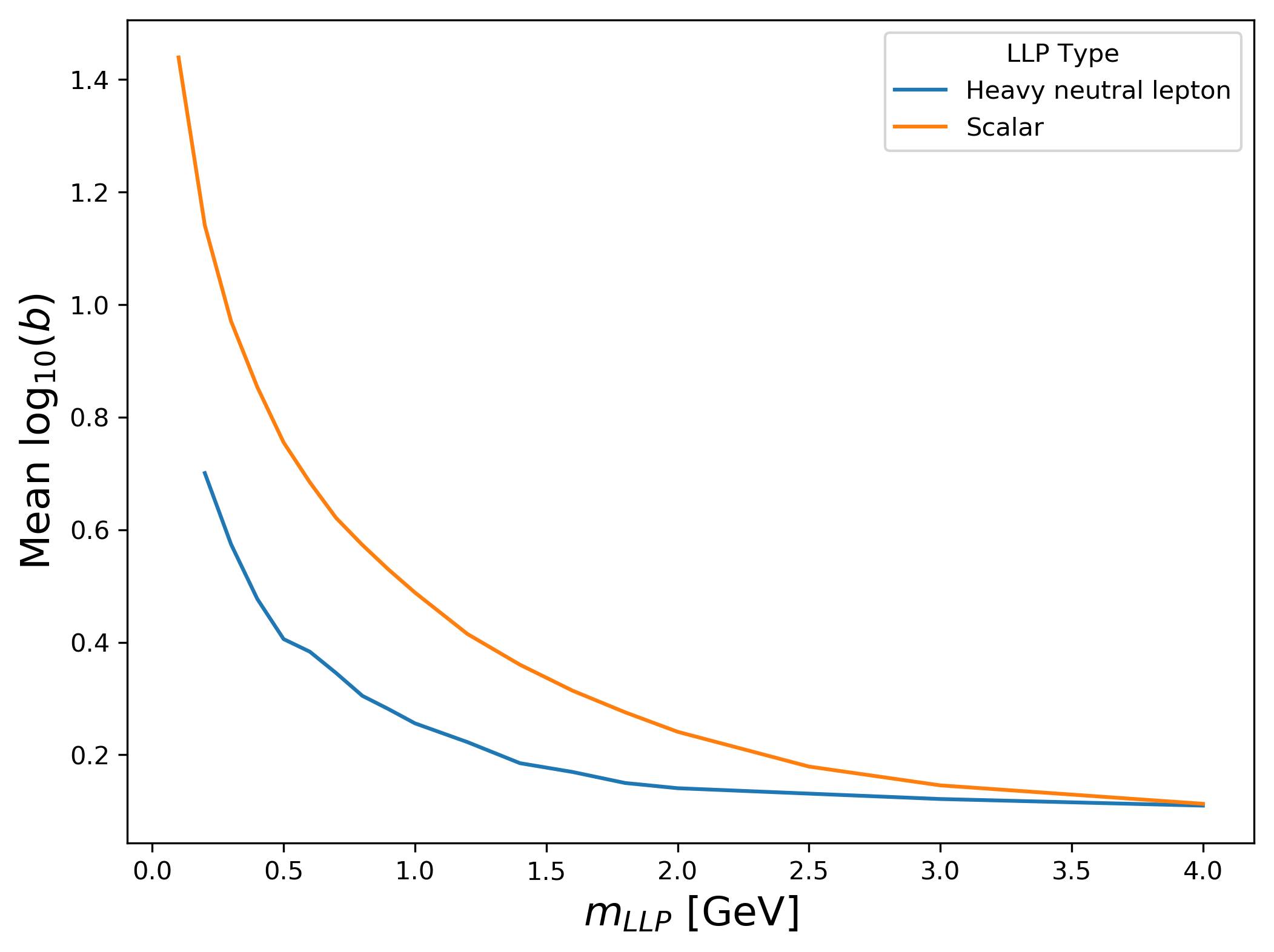}
 		\caption{Average LLP boost as a function of $m_{LLP}$ for production in $B$-meson decays, assuming the LLP is either a HNL or a light neutral scalar. }
 		\label{fig:BBboostmass}
 	\end{figure}

 	With the mass of SM mesons known, the only free parameter in this model is the LLP mass. If $m_{LLP} \ll m_{B}$, the LLPs produced will be highly boosted, and their boost is strongly correlated with their mass. Figure~\ref{fig:BBboostmass} shows this relationship for both scalar and RHN LLPs, using the mean boost to demonstrate the dependence on $m_{LLP}$. It should be noted that as $m_{LLP}$ increases, the boost rapidly loses its dependence on $m_{LLP}$. 
 	
 	The precision of the $m_{LLP}$ estimation was evaluated at three benchmark LLP masses: $0.5,\ 1.0,\ 3.0\ \GeV$, for both HNL and scalar LLPs. Maximum likelihood estimation was performed on many samples of $N_{obs}=10,100,1000$, comparing each sample's boost histogram against expected histograms for $\log_{10}(b_{LLP})$ at each mass point created from high-statistics simulated samples. Tables \ref{tab:BBparamest_HNL} and \ref{tab:BBparamest_scalar} show the standard deviation of best-fit masses for each mass point and pseudo-experiment sample size under the two different models. In cases where no local minimum in $-LLH$ was found within the range of masses with available simulation, typically occurring for high $m_{LLP}$, the best-fit LLP mass was set to 4 GeV. Consequently, the spread of best-fit masses is generally underestimated for the $m_{LLP}=3\ \GeV$ case. 
 	
 	Comparing the two kinds of LLP under consideration, we find that LLP mass can be estimated with lower standard error in the scalar case than the HNL case. At or below $\approx 1\ \GeV$, the HNL LLP mass can be determined with relative precision $\approx 0.15-0.2$ with $N_{obs}=100$, while the scalar LLP mass can be determined with relative precision $\approx 0.1$. At $m_{LLP}=3\ \GeV$, the spread in best-fit masses is large enough for both types of LLP that a large fraction of samples have no local $-LLH$ minimum within the range of masses simulated, indicating a large uncertainty in the LLP mass.     

 	\begin{table}
 		\centering
 	 	\hspace*{-8mm}
 		\begin{tabular}{|c||c|c|c||c|c|c||c|c|c|}
 			\hline
 			True $m_{LLP}$ [GeV] & \multicolumn{3}{c||}{0.5} & \multicolumn{3}{c||}{1.0} & \multicolumn{3}{c|}{3.0} \\
 			\hline
 			\hline
 			$N_{obs}$& 10 & 100 & 1000 & 10 & 100 & 1000 & 10 & 100 & 1000\\ \hline
 			$N_{samples}$ & 1083 & 110 & 10 & 1574 & 155 & 15 & 2144 & 213 & 21\\ \hline
		\begin{tabular}{c}Relative precision
	of \\ best-fit $m_{LLP}$ \end{tabular} & 0.68 & 0.17 & 0.051 & 0.71 & 0.19 & 0.047 & > 0.43 &$\gtrapprox$  0.28 & $\gtrapprox$ 0.17 \\
 			\hline
 		\end{tabular}
 		\caption{Exotic B decay $\rightarrow $ Heavy neutral lepton LLP + SM mass estimation results. $N_{samples}$ is the number of pseudodata sets used to estimate the spread of best-fit values. Relative spread of best-fit masses is reported for three choices of $m_{LLP}$ and $N_{obs}=10,100,1000$.}
 		\label{tab:BBparamest_HNL}	
 	\end{table}
 	
 	\begin{table}
	 	\centering
	 	\hspace*{-10mm}
	 	\begin{tabular}{|c||c|c|c||c|c|c||c|c|c|}
	 		\hline
	 		True $m_{LLP}$ [GeV] & \multicolumn{3}{c||}{0.5} & \multicolumn{3}{c||}{1.0} & \multicolumn{3}{c|}{3.0} \\
	 		\hline
	 		\hline
	 		$N_{obs}$& 10 & 100 & 1000 & 10 & 100 & 1000 & 10 & 100 & 1000\\ \hline
	 		$N_{samples}$ & 504 & 50 & 5 & 946 & 95 & 9 & 1993 & 200 & 20 \\ \hline
		\begin{tabular}{c}Relative precision
	of \\ best-fit $m_{LLP}$ \end{tabular} & 0.29 & 0.092 & 0.020 & 0.34 & 0.091& 0.032 & > 0.28 & $\gtrapprox$ 0.16 & $\gtrapprox$ 0.052 \\
	 		\hline
	 	\end{tabular}
	 	\caption{Exotic B decay $\rightarrow $ Scalar LLP + SM mass estimation results. $N_{samples}$ is the number of pseudodata sets used to estimate the spread of best-fit values. Relative spread of best-fit masses is reported for three choices of $m_{LLP}$ and $N_{obs}=10,100,1000$.}
	 	\label{tab:BBparamest_scalar}
 \end{table}

\subsection{Charged Current}

Second, let us consider the Charged Current production model, where two new particles participate in the LLP production process. We need to define a variable that is complementary to the LLP boost $b = x_1$, in order to determine both parent and LLP masses. 
    
In the CC model, the LLP is produced in a decay whose other product is a SM lepton $\ell$, which is most likely observed and whose momentum is measured in CMS. Its transverse momentum $p_{T}^{\ell}$ is highly correlated with $m_{parent}-m_{LLP}$, which is an orthogonal combination of masses to $m_{LLP}$. We therefore set $x_2 = p_T^\ell$.\footnote{We also checked the performance of an estimator that uses the full lepton momentum measured at CMS and the LLP velocity measured at MATHUSLA   to directly reconstruct the parent mass, but surprisingly we found that using just $p_{T}^{\ell}$ gave better precision, likely because it depends more directly on $m_{parent} - m_{LLP}$.} 
   
\begin{figure}
	\begin{center}
	\begin{tabular}{cc}
	 		\includegraphics[width=.45\textwidth]{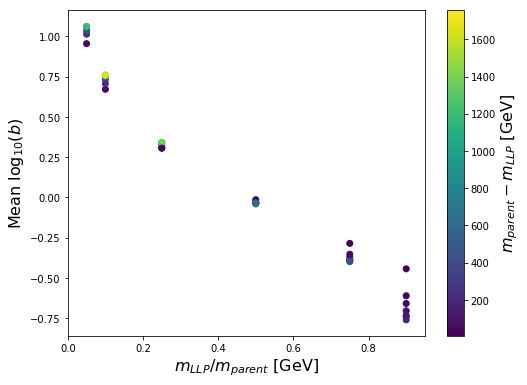}
			&
			\includegraphics[width=.45\textwidth]{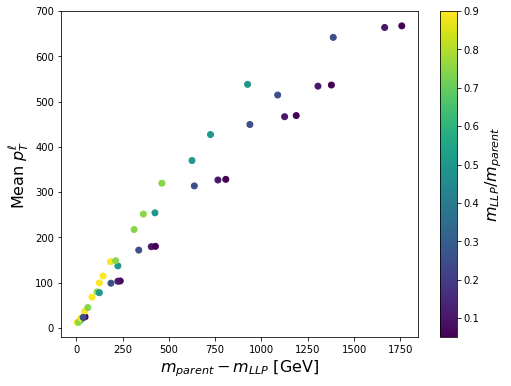}
		\end{tabular}
		\end{center}
		\vspace*{-5mm}
 		\caption{
		For the Charged Current (CC) production mode, we show on the left and right plots how the high-statistics distributions of our estimator variables $x_1 = \log_{10}(b)$ and $x_2=p_{T}^{\ell}$ depend on $m_{LLP}/m_{parent}$ and $m_{parent}$ in complementary ways. The color legend shows the subdominant variation with the complementary combination of LLP and parent mass.}
  		\label{fig:CCestimators}
 	\end{figure}

The strong dependences of the high-statistics distribution average of $x_{1}=\log_{10}(b)$ on $m_{LLP}/m_{parent}$, and the peak value of the $x_2=p_{T}^{\ell}$ distribution on $m_{parent}-m_{LLP}$ are illustrated in Figure~\ref{fig:CCestimators}.

	\begin{figure}
	\centering
	\includegraphics[width=.6\textwidth]{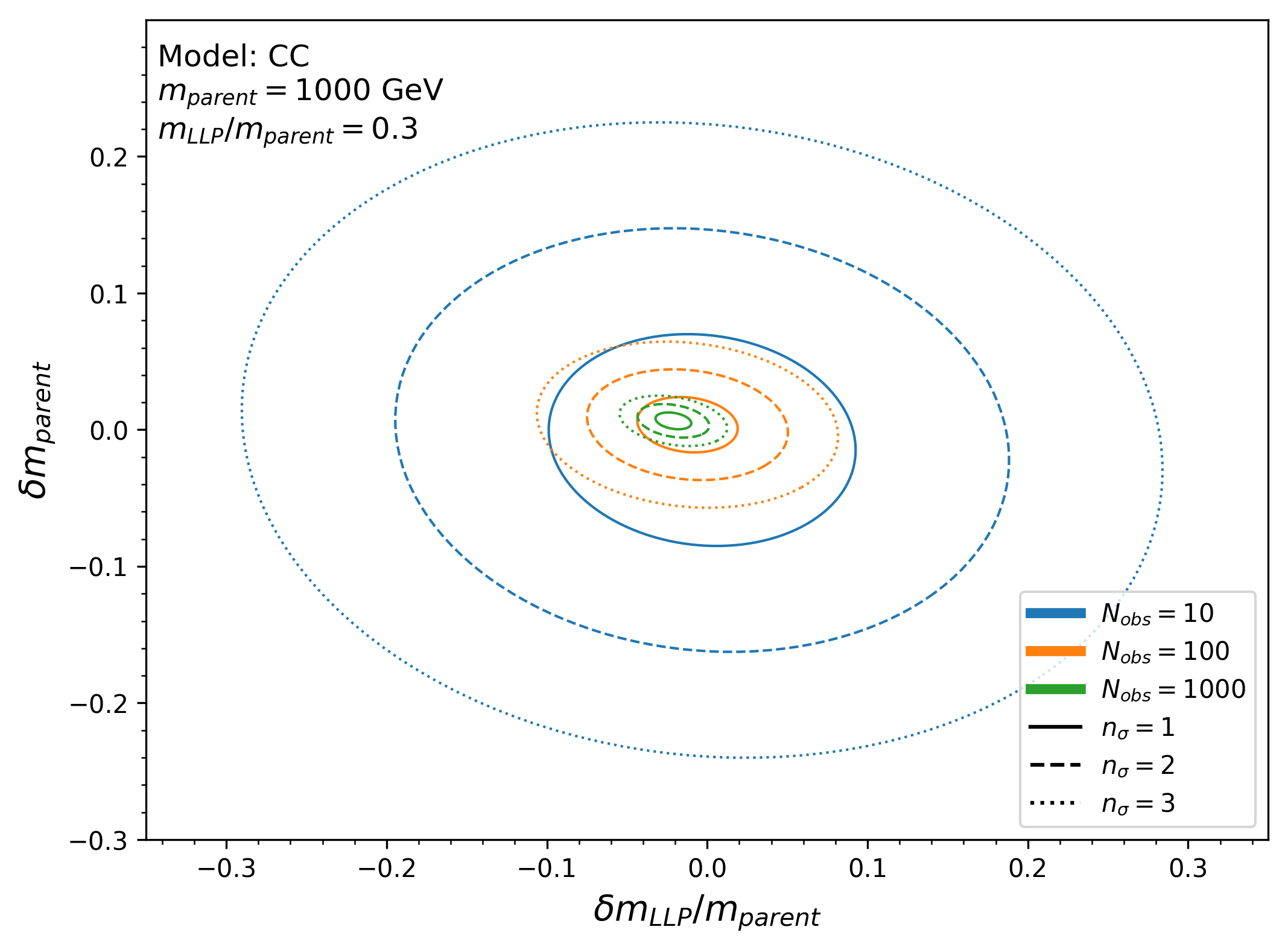}
	\caption{Parameter estimation for a benchmark point $(m_{parent}, m_{LLP}/m_{parent}) = (1000\ \GeV, 0.3)$ in the Charged Current model, showing 1, 2, and 3$\sigma$ ellipses for $N_{obs}=10,100,1000$ observed events. The axes are in units of $\delta x \equiv \frac{x_{fit}-x_{true}}{x_{true}}$, where $x_{true}$ and $x_{fit}$ are the actual values of the parameters and their measured value extracted from the fit. 
	}
	\label{fig:CCellipses}
	\end{figure} 
 
\begin{table}
	\centering
	
	\begin{tabular}{|l|c|c|c|}
		\hline
		$N_{obs}$& 10 & 100 & 1000 \\
		$N_{samples}$ &838 &84 &8 \\
		Relative precision of $m_{LLP}/m_{parent}$ &0.096 &0.032 &0.011 \\
		Relative precision of $m_{parent}$ &0.078 &0.020 &5.7$\times 10^{-3}$ \\
		Pearson correlation coefficient & -0.09& -0.14&-0.33 \\
		\hline
	\end{tabular}
	\caption{Charged Current model mass estimation results with $m_{LLP}=300\ \GeV$, $m_{parent} = 1\ \TeV$. $N_{samples}$ is the number of pseudo-data sets tested for the indicated number of events $N_{obs}$. The low correlation coefficient indicates that the uncertainties in $m_{LLP}/m_{parent}$ and $m_{parent}$ separately are not highly correlated.}
	\label{tab:CCbfmassestimationresults}
\end{table}

Figure~\ref{fig:CCellipses} shows the precision attained for $m_{parent}$ and $m_{LLP}/m_{parent}$ in the cases $N_{obs}=10,100,1000$ 	for our benchmark parameter point of $(m_{parent}, m_{LLP}) = $ (1 TeV, 300 GeV). Table \ref{tab:CCbfmassestimationresults} gives the relative precisions for parent and LLP masses, defined as the standard deviation divided by the mean of best-fit masses. 
 The most important result is that under the CC production model, with 100 LLP observations in MATHUSLA successfully matched with their production events in CMS, the parent particle mass and the $m_{LLP}/m_{parent}$ mass ratio can both be determined with better than 5\% resolution. Even with only 10 events, these resolutions degrade to $\approx 8\%$ and 10\% respectively.

\subsection{Heavy Parent}

\begin{figure}
 		\begin{center}
 		\begin{tabular}{cc}
	 		\includegraphics[width=.45\textwidth]{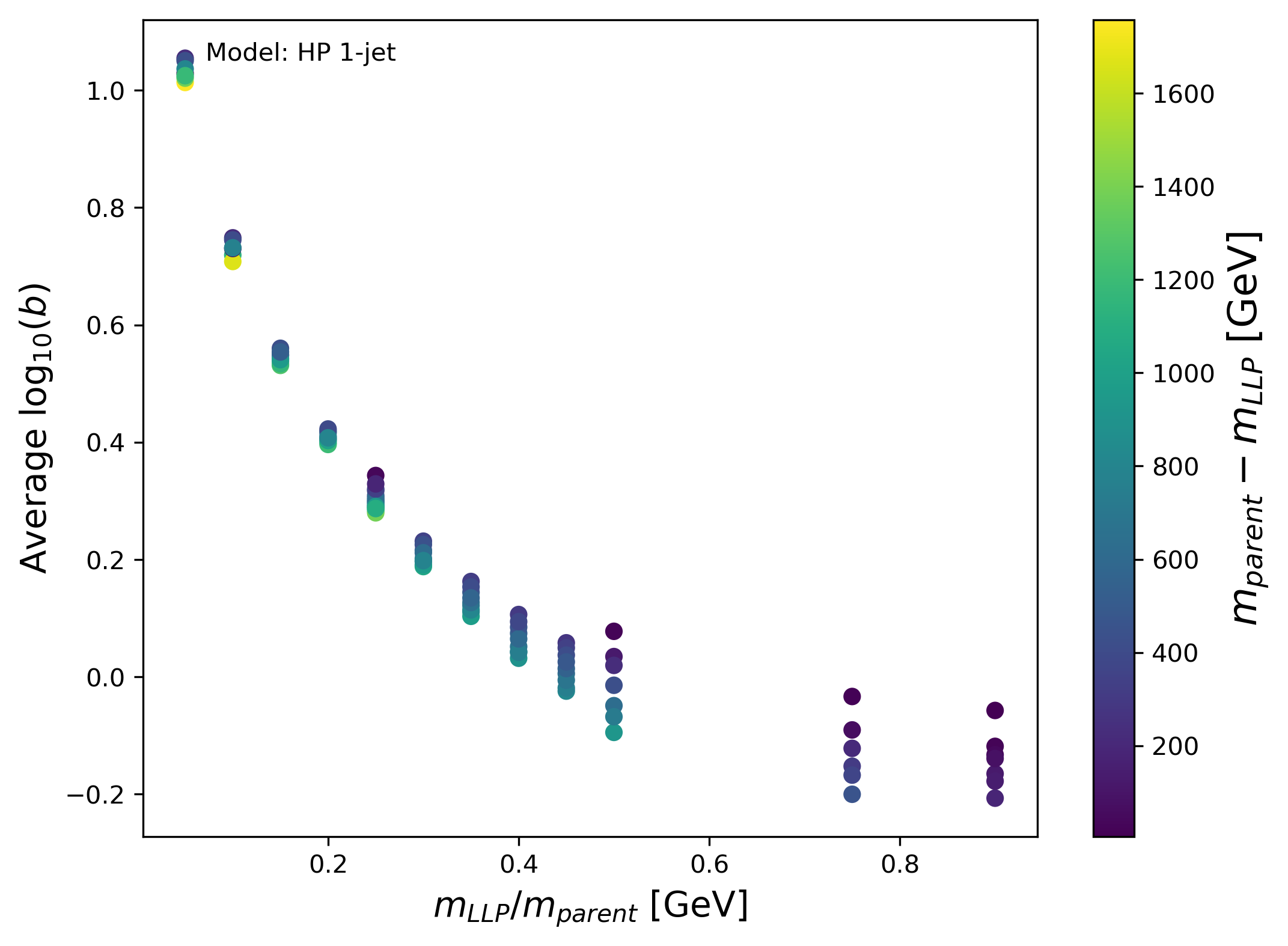}
	 	&
			\includegraphics[width=.45\textwidth]{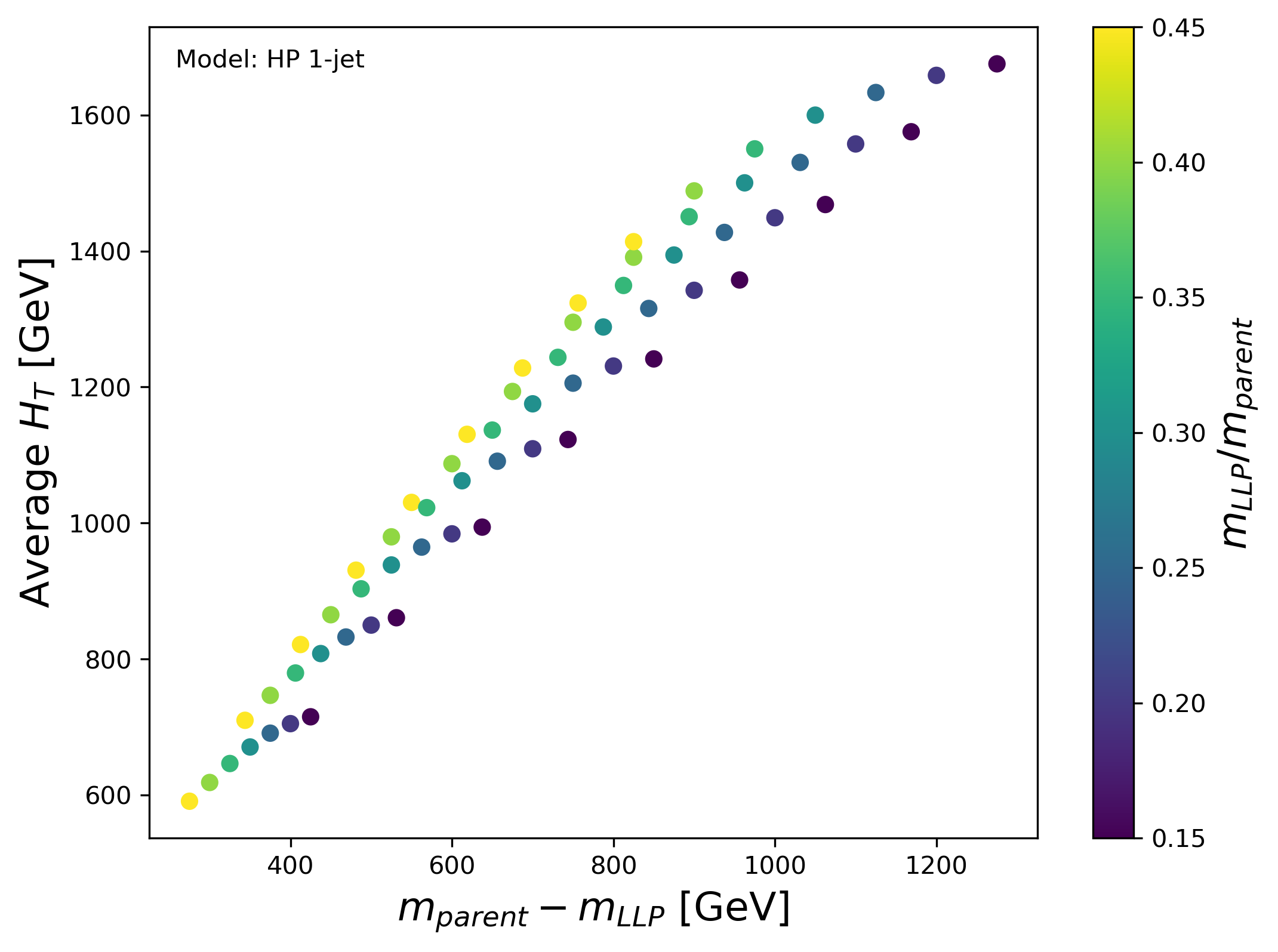}
 		\end{tabular}
 		\end{center}
 		\caption{For the Heavy Parent (HP) production mode, we show on the left and right plots how the high-statistics distributions of our estimator variables $x_1 = \log_{10}(b)$ and $x_2=H_{T}$ depend on $m_{LLP}/m_{parent}$ and $m_{parent} - m_{LLP}$ in complementary ways. The color legend shows the subdominant variation with the complementary combination of LLP and parent mass.}
 		\label{fig:HPestimators}
 	\end{figure}

Next, consider the Heavy Parent model. Again, one of our variables will be the LLP boost $\log_{10}(b) = x_1$, and we need to find a complementary variable $x_2$. We select $x_{2} = H_{T}$, the scalar sum of jet transverse momenta in CMS. Its distribution is highly correlated with $m_{parent}-m_{LLP}$, similarly to $p_{T}^{\ell}$ in the Charged Current case, and works well for both one- and two-jet parent decay scenarios.\footnote{Again, as for the Charged Current mode, we investigated variables that directly reconstruct the parent mass using MATHUSLA and CMS information, which in this case have to also contend with combinatorics background. In analogy to the CC analysis, the simple $H_T$ again delivered superior performance.} 
Figure~\ref{fig:HPestimators} shows the dependence of the high-statistics average value of both estimators on the BSM particle masses. 

 	\begin{figure}
 		\begin{center}
 		 \begin{tabular}{cc}
 			\centering
 			\includegraphics[width=.45\textwidth]{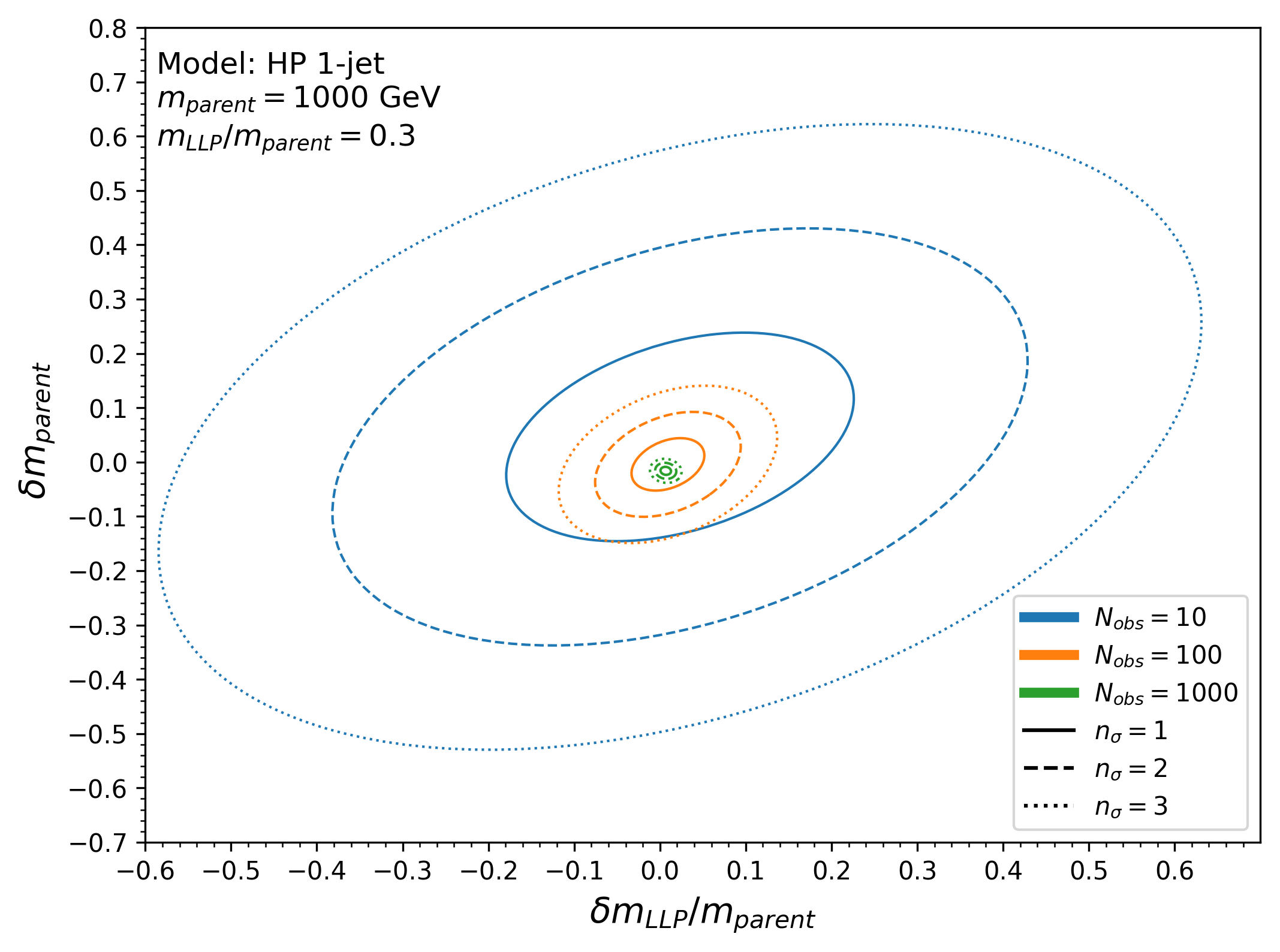}
 		&
			\includegraphics[width=.45\textwidth]{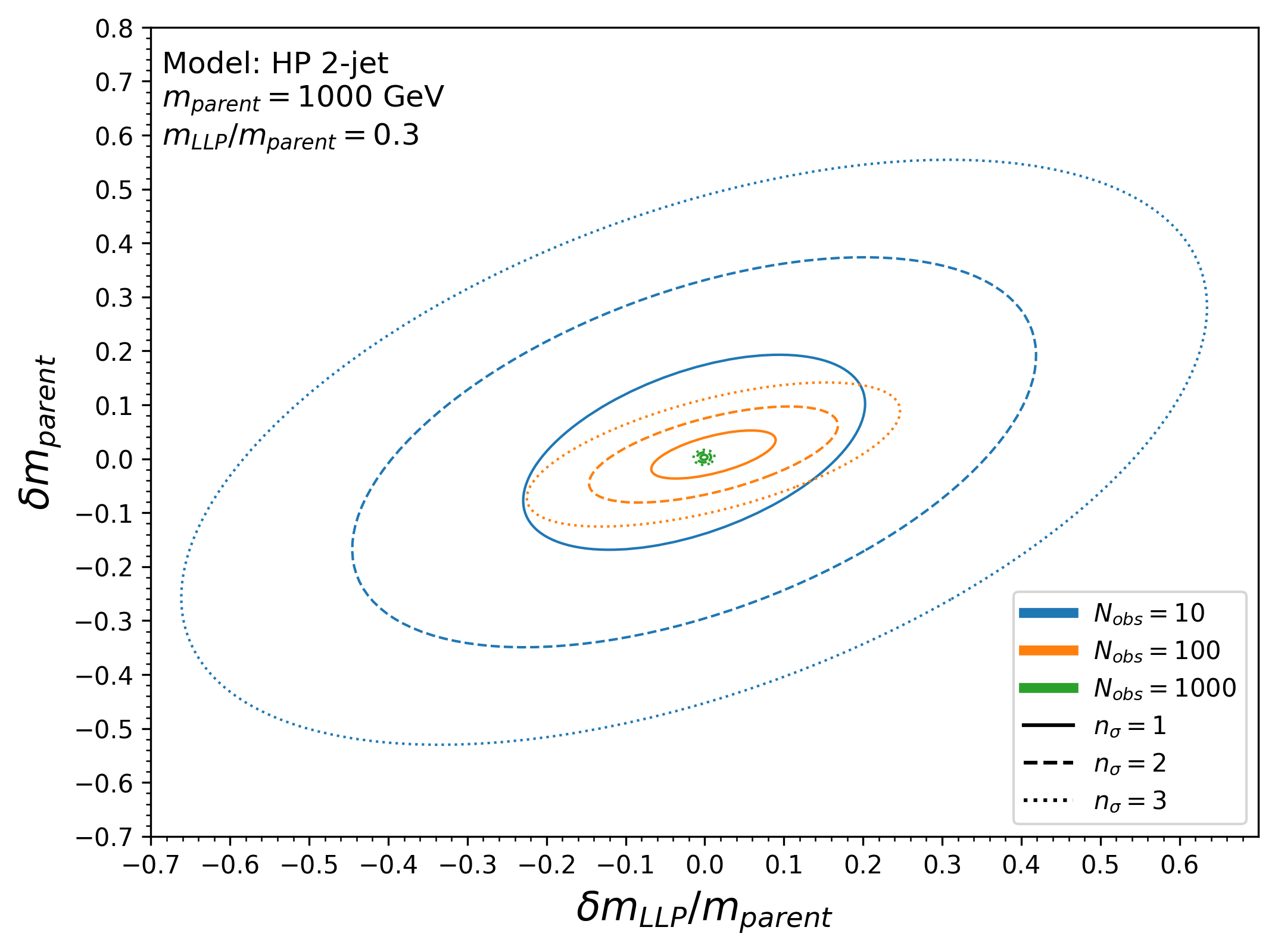}	
	 	\end{tabular}
 		\end{center}
 		\caption{Parameter estimation results for a benchmark point $(m_{parent}, m_{LLP}/m_{parent}) = (1000\ \GeV, 0.3)$ in the Heavy Parent model, with 1-jet decay on the left and 2-jet decay on the right, showing 1, 2, and 3$\sigma$ ellipses for $N_{obs}=10,100,1000$ observed events. The axes are in units of $\delta x \equiv \frac{x_{fit}-x_{true}}{x_{true}}$, where $x_{true}$ and $x_{fit}$ are the actual values of the parameters and their measured value extracted from the fit. }
 		\label{fig:HPHTellipses} 
 		
    \end{figure}

	 Ellipses showing the spread of best-fit masses for pseudo-data samples of size  $N_{obs}=10,100,1000$ using $H_{T}$ are displayed in Figure~\ref{fig:HPHTellipses} for the one- and two-jet decay modes of our  benchmark parameter point $(m_{parent}, m_{LLP}) = $ (1 TeV, 300 GeV). 
	 Tables $\ref{tab:HPbfmassestimationresults}$ and $\ref{tab:HP2Jbfmassestimationresults}$ summarize the distributions of best-fit particle masses in the one- and two-jet Heavy Parent decay models respectively. For the one-jet case, 5\% resolution in $m_{parent}$ and $m_{LLP}/m_{parent}$ is possible with 100 observed LLP events in MATHUSLA, and 20\% resolution with just 10 events. In the two-jet case, resolution of 5-10\% in $m_{parent}$ and $m_{LLP}/m_{parent}$ is possible with 100 observed LLP events in MATHUSLA, and 20-30\% resolution with just 10 events. 

  	\begin{table}
	 	\centering

	 	\begin{tabular}{|l|c|c|c|}
	 		\hline
	 		$N_{obs}$& 10 & 100 & 1000 \\
	 		$N_{samples}$ & 1154 &116 & 11\\
	 		Relative precision of $m_{LLP}/m_{parent}$ &0.20& 0.042& 5.8$\times 10^{-3}$\\
	 		Relative precision of $m_{parent}$ [GeV]& 0.18&0.048&7.2$\times 10^{-3}$\\
	 		Pearson correlation coefficient & 0.37&0.34&-0.093 \\
	 		\hline
	 	\end{tabular}
 		\caption{Heavy Parent 1-jet decay model mass estimation results for $m_{LLP}=300\ \GeV$, $m_{parent} = 1\ \TeV$, using estimators $x_1 = \log_{10}(b_{LLP}),\  x_2 = H_{T}$. $N_{samples}$ is the number of pseudo-data sets tested for the indicated number of events $N_{obs}$.}
	 	\label{tab:HPbfmassestimationresults}	
	 \end{table}
  	\begin{table}
	\centering
	\begin{tabular}{|l|c|c|c|}
		\hline
		$N_{obs}$& 10 & 100 & 1000 \\
		$N_{samples}$ & 2450& 244&23\\
		Relative precision of $m_{LLP}/m_{parent}$ & $\gtrapprox$ 0.22&0.078&4.3$\times 10^{-3}$ \\
		Relative precision of $m_{parent}$ [GeV]& 0.18&0.044&4.5$\times 10^{-3}$ \\
		Pearson correlation coefficient & 0.50&0.61&0.054 \\
		\hline
	\end{tabular}
	\caption{Heavy Parent 2-jet decay model mass estimation results for $m_{LLP}=300\ \GeV$, $m_{parent} = 1\ \TeV$, using estimators $x_1 = \log_{10}(b_{LLP}),\  x_2 = H_{T}$. $N_{samples}$ is the number of pseudo-data sets tested for the indicated number of events $N_{obs}$.}
	\label{tab:HP2Jbfmassestimationresults}	
	\end{table}

Since the $H_T$ variable can be used to estimate masses in the one-jet and two-jet Heavy Parent decay scenarios, it is interesting to consider whether the parameter estimation even requires an accurate determination of the production mode (within the Heavy Parent class of scenarios). In other words, could one measure the parent particle mass in the HP scenario, without knowing the exact hadronic decay mode of the parent? 
To test this, we performed a parameter estimation on HP two-jet data with $m_{parent}=1000\ \GeV$, $m_{LLP}=300\ \GeV$, but using the one-jet hypothesis as our fitting templates. 
The best-fit masses had a large positive bias, in many cases beyond the range of simulated masses. This is not unexpected, since the final states of a 2-body decay have lower  combined $H_T$ than the final states of a 3-body decay. This bias should therefore persist for lower masses, indicating that precise measurements of $m_{parent}$ and $m_{LLP}$ requires fairly precise knowledge of the parent decay mode.\footnote{It might be possible to construct some $H_T$-like variable that takes final state jet multiplicity into account event-by-event, such that a measurement of $m_{parent}$ becomes possible even without a detailed hypothesis on the precise decay mode. We leave this for future investigations.}

\subsection{Exotic Higgs Decay}
 	
\begin{figure}
\begin{center}
\begin{tabular}{cc}
\includegraphics[width=.45\textwidth]{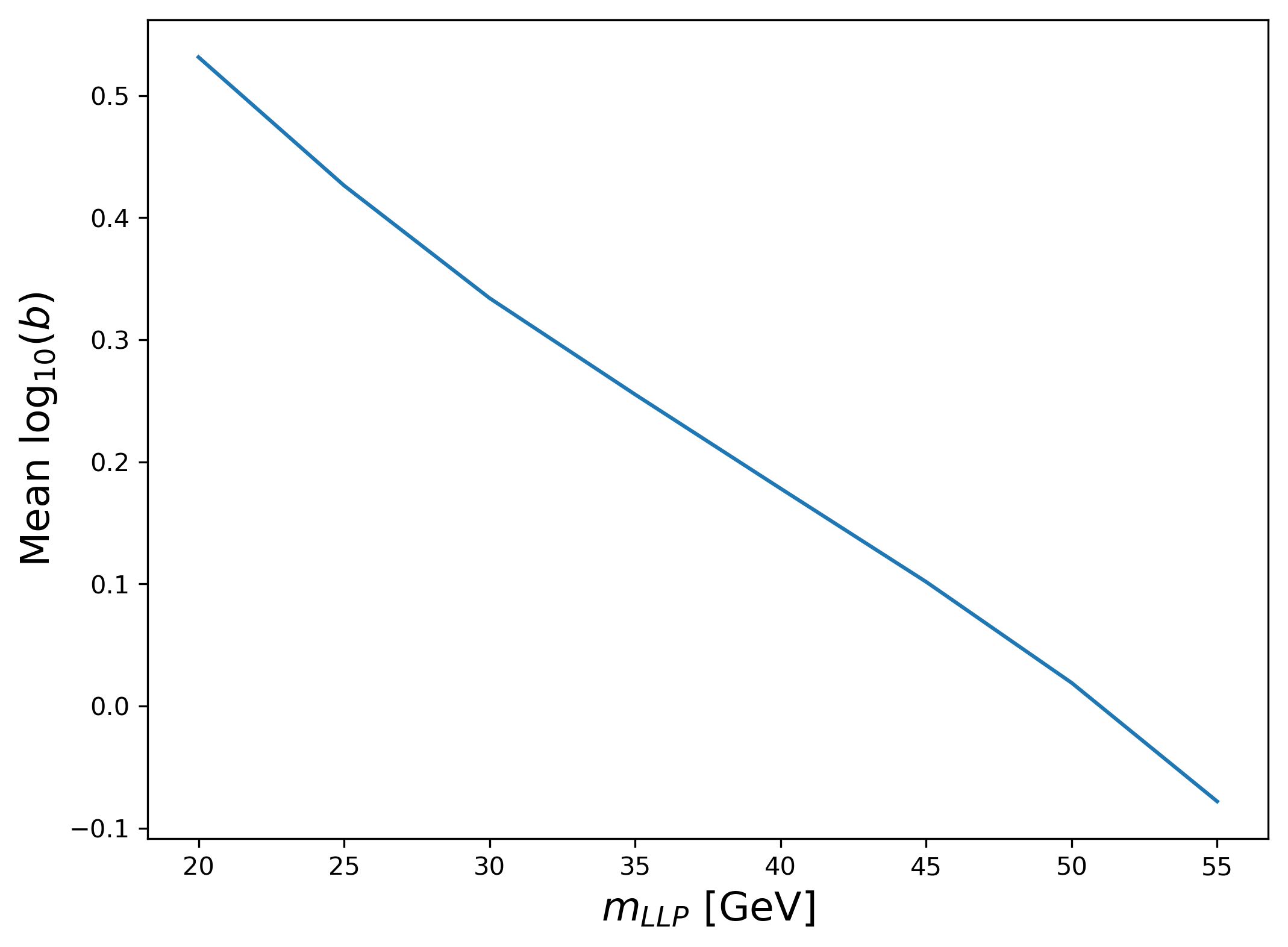}
&
\includegraphics[width=.45\textwidth]{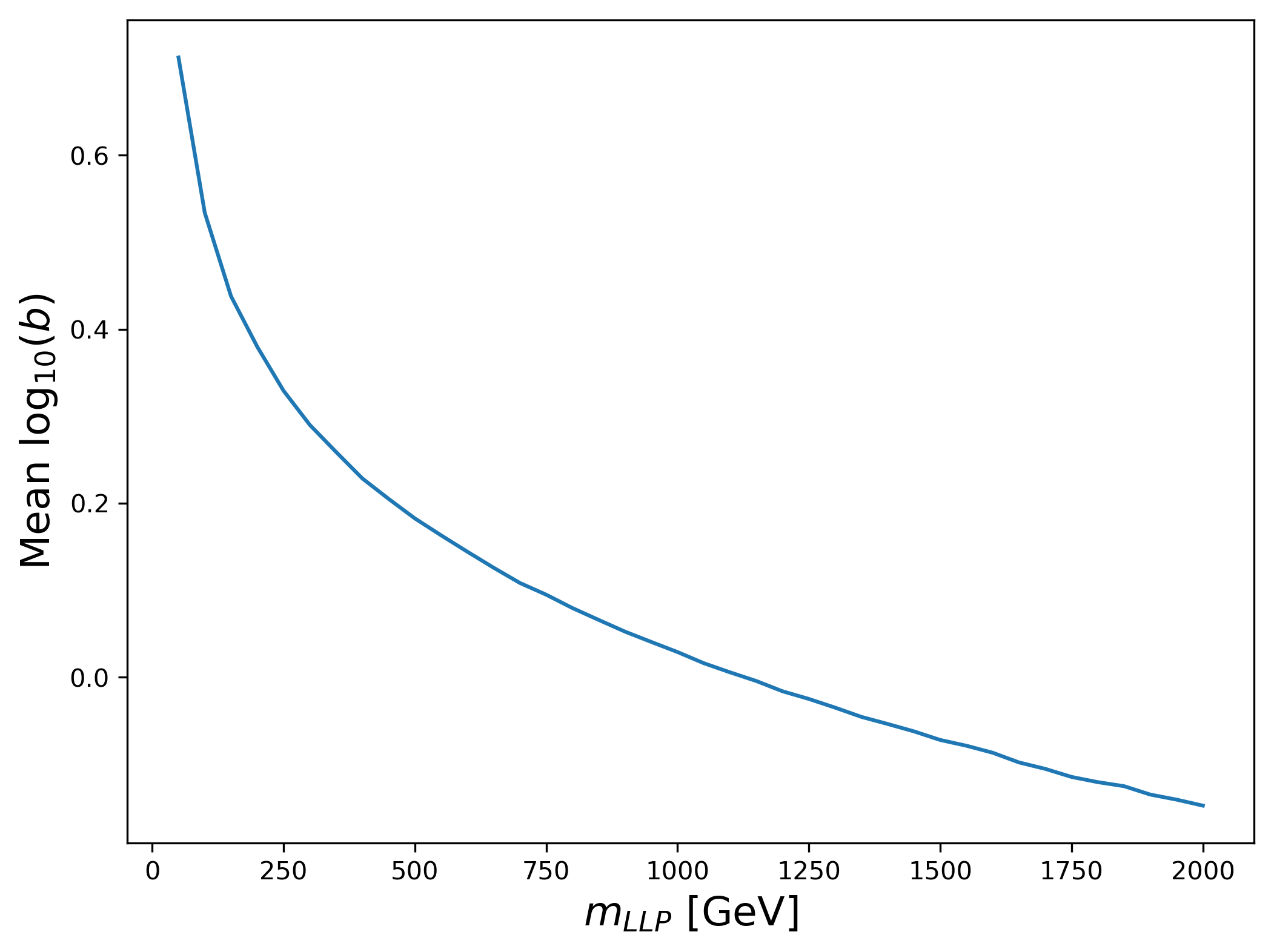}
\end{tabular}
\end{center}
\vspace*{-5mm}
\caption{
Left: average LLP boost as a function of LLP mass in the Exotic Higgs Decay model. Right: For the Direct Pair Production model. Note the dependence becomes weaker at higher masses for DPP.
}
\label{fig:boostvsLLPmassHIGDPP}
\end{figure}

In the Exotic Higgs Decay model, the average boost $b\sim m_h/2 m_{LLP}$ of LLPs is strongly correlated with LLP mass, see Figure~\ref{fig:boostvsLLPmassHIGDPP} (left). This scenario was already analyzed in Ref.~\cite{Curtin:2017izq}, but we include it here for completeness. Maximum likelihood estimation was performed on the boost distributions of many pseudo-data samples at $m_{LLP}=35\ \GeV$, similarly to the procedure for exotic $B$-meson decays to evaluate the mass estimation precision. Because the precision does not vary significantly over the range of $m_{LLP}$ considered for this model, we only show results  for this one mass point. Table \ref{tab:HIGbfmassestimationresults} summarizes the distributions of best-fit masses for $N_{obs}=10,100,1000$.

	\begin{table}
		\centering
		\begin{tabular}{|c|c|c|c|}
			\hline
			$N_{obs}$& 10 & 100 & 1000 \\
			$N_{samples}$ & 3229 & 321 & 32\\
			Relative precision of $m_{LLP}$ & 0.15 & 0.037 & 0.011\\
			\hline
		\end{tabular}
		\caption{Exotic Higgs decay model mass estimation results with estimator $\log_{10}(b_{LLP})$ and true $m_{LLP}=35\ \GeV$. $N_{samples}$ is the number of pseudo-data sets tested for the indicated number of events $N_{obs}$.}
		\label{tab:HIGbfmassestimationresults}	
	\end{table}

\subsection{Direct Pair Production}
Our strategy for the Direct Pair Production model is the same as for Higgs decay. The correlation between boost and LLP mass is strong for this model, degrading in the high-mass limit, see Figure~\ref{fig:boostvsLLPmassHIGDPP} (right). Maximum likelihood estimation was performed for ensembles of pseudo-experiments with $N_{obs}=10,100,1000\ \GeV$, for benchmark masses $100,1000,1700\ \GeV$. Table \ref{tab:DPPbfmassestimationresults} gives the mean and relative precision of the best-fit masses obtained for each choice. Note that since the highest mass point simulated was 2 TeV, this is the upper bound on best-fit masses reported. Therefore, the spread in best-fit masses may be underestimated for $m_{LLP}=1700\ \GeV$, $N_{obs}=10,100$ (although the prior on detecting much heavier states at the LHC would be suppressed due to kinematic suppression of the production rate). For all three benchmark masses, resolution of $15\%$ or better can be achieved with 100 observed LLPs.

     \begin{table}
    	\centering
    	\hspace*{-12mm}
    	\begin{tabular}{|c||c|c|c||c|c|c||c|c|c|}
    		\hline
    		True $m_{LLP}$ [GeV] & \multicolumn{3}{c||}{100} & \multicolumn{3}{c||}{1000} & \multicolumn{3}{c|}{1700} \\
    		\hline
    		\hline
    		$N_{obs}$& 10 & 100 & 1000 & 10 & 100 & 1000 & 10 & 100 & 1000\\
    		$N_{samples}$ &779 &78 & 8 & 2523 & 244& 25& 2372 &237 &23 \\
    		Relative precision of $m_{LLP}$ & 0.50 & 0.13 & 0.033 & .28 & 0.085 & 0.049 & > 0.25 & $\gtrapprox$ 0.14 & 0.068 \\
    		\hline
    	\end{tabular}
   		\caption{Direct Pair Production model mass estimation results with estimator $\log_{10}(b_{LLP})$. $N_{samples}$ is the number of pseudo-data sets tested for the indicated number of events $N_{obs}$.}
    	\label{tab:DPPbfmassestimationresults}	
    \end{table}

\subsection{Heavy Resonance}
 	\label{subsection:resparamestimation}

The final simplified model to consider is the intermediate $s$-channel Heavy Resonance. Like the Heavy Parent scenario, there are two sub-categories. In this case they are the narrow- and high-width limits for the resonant particle. Just as for HP, we will first consider parameter estimation in each case separately, then investigate the impact of using the incorrect width hypothesis to estimate parameters. Finally, we will demonstrate that the shape of the boost distribution provides useful information on the resonance width. 
	
\begin{figure}
 		\begin{center}
 			\begin{tabular}{cc}
 			\includegraphics[width=0.45\textwidth]{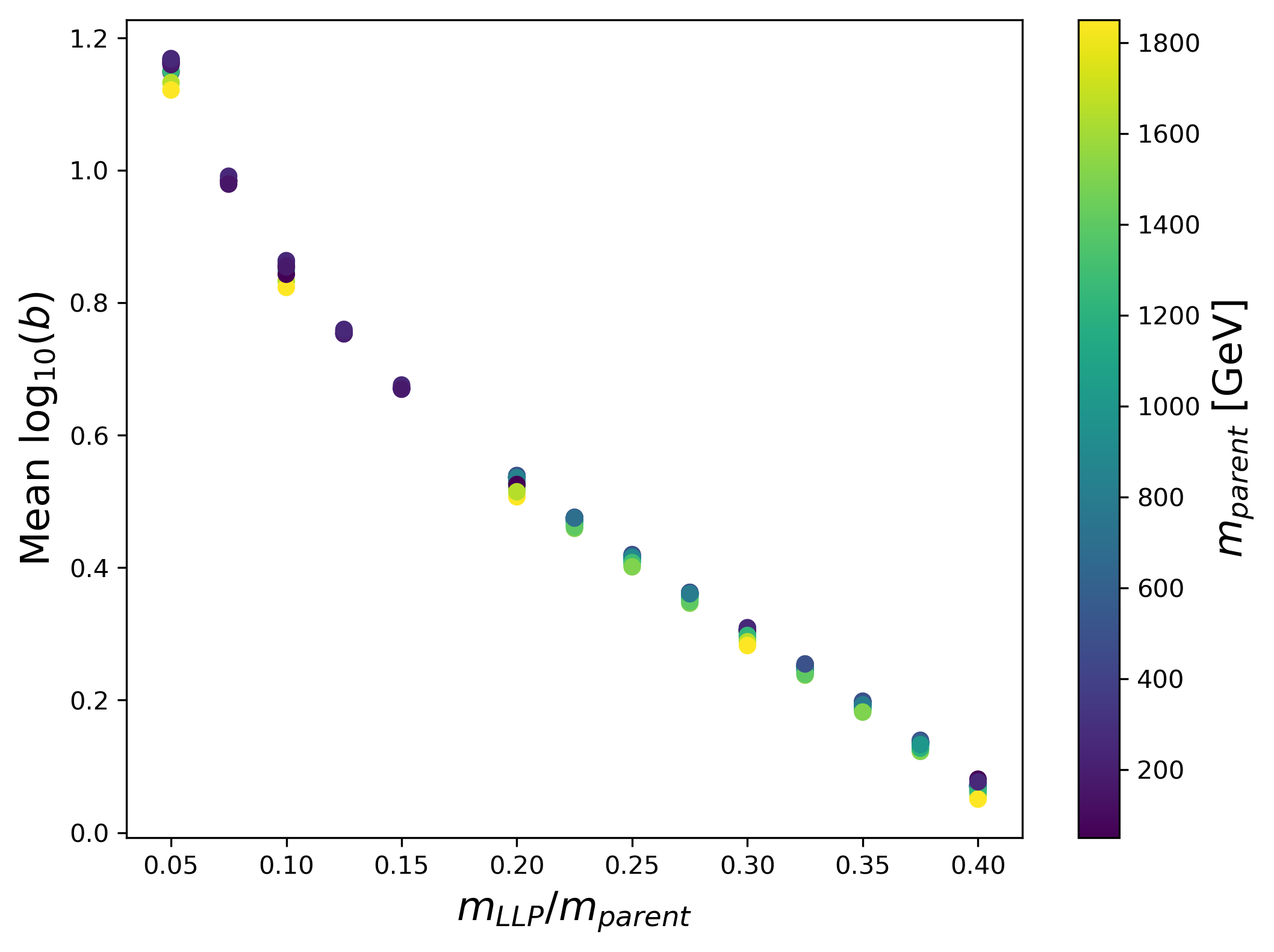}
 			&
 			\includegraphics[width=.45\textwidth]{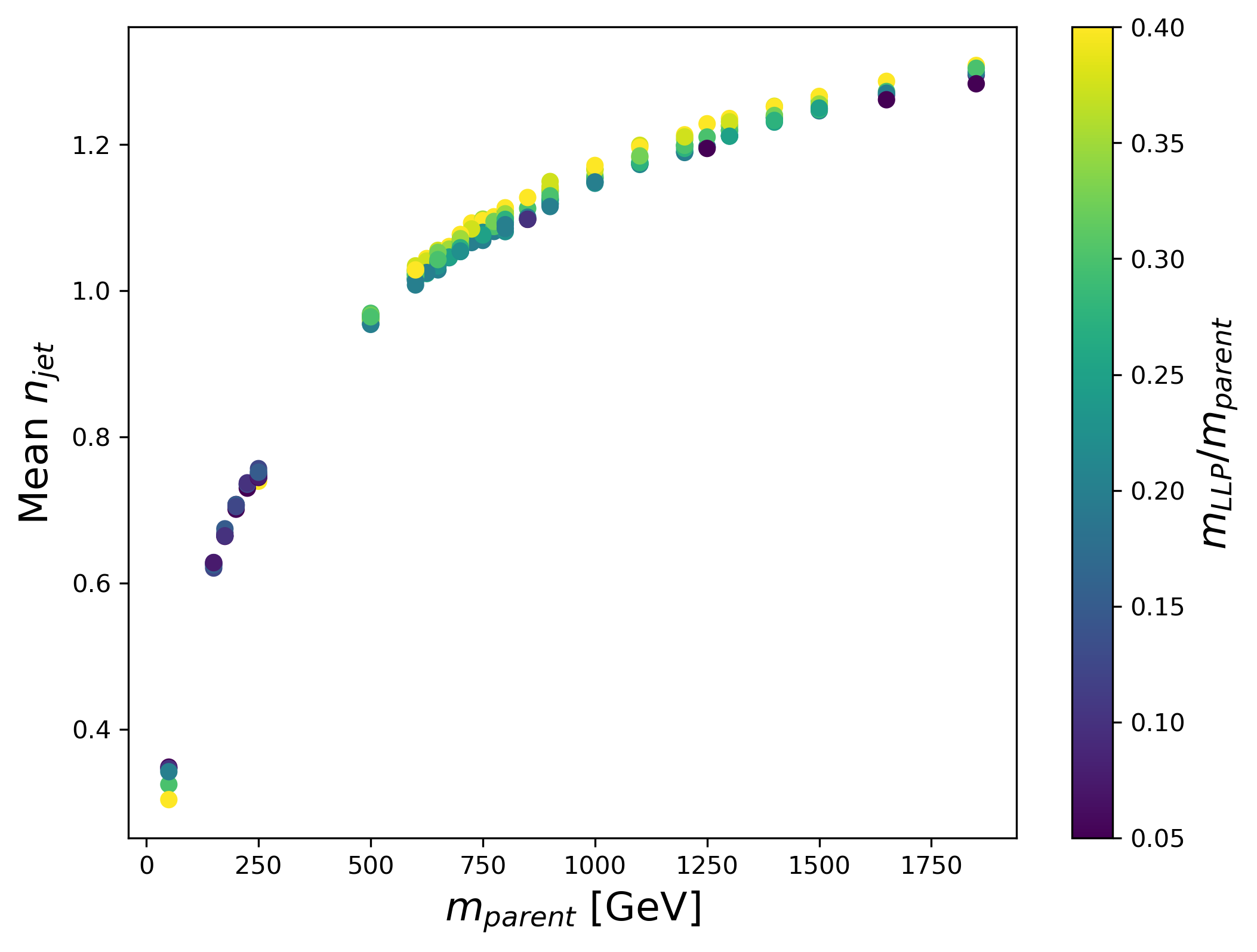}
 			\end{tabular}
 		\end{center}
 		\caption{For the Heavy Resonance (RES) production mode, we show on the left and right plots how the high-statistics distributions of our estimator variables $x_1 = \log_{10}(b)$ and $x_2=n_{jet}$ depend on $m_{LLP}/m_{parent}$ and $m_{parent}$ in complementary ways. The color legend shows the subdominant variation with the complementary combination of LLP and parent mass.}
 		\label{fig:RESestimators}
 	\end{figure}

In the RES simplified model, the LLP boost distribution is almost entirely dependent on $m_{LLP}/m_{parent}$, with very little separate dependence on $m_{parent}$. This is demonstrated in Figure~\ref{fig:RESestimators} (left). This not only makes $x_1 = b$ the obvious choice for one of our fit variables, it also means the LLP to parent mass ratio can be extracted independently.
	
To find the second complementary variable that independently measures $m_{parent}$, the only handle we have available is the  fact that the energy scale of ISR jets in the detector is determined by the overall energy scale of the hard process, which is $m_{parent}$. Several variables were considered as candidates for $x_2$, including the highest main-detector jet transverse momentum, $p_{T}^{j1}$, the scalar sum of jet transverse momenta $H_{T}$, and the number of jets with $p_{T}>20\ \GeV$, $n_{jet}$. It was found that $n_{jet}$ gave the best sensitivity, and its dependence on $m_{parent}$ is shown in Figure~\ref{fig:RESestimators} (right). The weakening dependence of $\langle n_{jet}\rangle$ at high $m_{parent}$ is consistent with the logarithmic dependence of ISR kinematics on the hard scale of the event. 
	
These two variables were used for the two-dimensional binned likelihood fit to measure $m_{parent}$ and $m_{LLP}/m_{parent}$ in the same fashion as the Charged Current and Heavy Parent analyses, using pseudo-data samples with $m_{parent}=1000\ \GeV$, $m_{LLP}=300\ \GeV$. Unfortunately, the variance in $n_{jet}$ is so high that only for $N_{obs}=1000$ was $m_{parent}$ estimated with precision better than the range of heavy resonance masses simulated. We therefore do not report a precision for measurement of $m_{parent}$ for $N_{obs} \lesssim 100$ for our benchmark point, but given the steeper $m_{parent}$ dependence of $n_{jet}$ for lighter parent masses, we expect that observation of fewer events might be sufficient to measure parent masses in the $\mathcal{O}(100 ~\mathrm{GeV})$ range.

Parameter estimation in the Heavy Resonance model is likely to benefit from other measurements at the LHC main detectors, since the $s$-channel mediator must have a visible SM decay channel, at minimum through the same process by which it was produced. This is similar to dark matter simplified models involving a heavy vector mediator, where some of the strongest constraints come from LHC searches for di-jet resonances \cite{Fairbairn:2016iuf,Chala:2015ama,Alexander:2016aln,Bagnaschi:2019djj}. Current and future colliders have the potential to detect the heavy mediator and measure its mass with $\mathcal{O}(10\%)$ or better precision with standard resonance search techniques \cite{Khachatryan:2010jd,Sirunyan:2019vxa}. If LLPs are discovered at MATHUSLA and classified as being produced via an $s$-channel resonance, the observed events in MATHUSLA could be used to independently measure $m_{LLP}/m_{parent}$. Table~\ref{tab:RESbfmassestimationresults} demonstrates that this ratio can be measured with $\sim 5\%$ precision or better for 10-100 observed events. Resonance searches at the main detectors would then provide an independent measurement of $m_{parent}$, allowing all mass parameters of the production mode to be determined. 
 	
  	\begin{table}
	\centering
	\begin{tabular}{|c|c|c|c|}
		\hline
		$N_{obs}$& 10 & 100 & 1000 \\
		$N_{samples}$ & 6565 (5104)&2279 (508) &228 (51) \\
		Relative precision of $m_{LLP}/m_{parent}$ &0.066 (0.12)&0.022 (0.044)&7.2$\times 10^{-3}$ (0.017)\\
		Relative precision of $m_{parent}$ [GeV]&- &-&0.12 (0.12)\\
		Pearson correlation coefficient &-&-& -0.57 (-0.84)\\
		\hline
	\end{tabular}
	\caption{Heavy Resonance model mass estimation results for $m_{LLP}=300\ \GeV$, $m_{parent} = 1\ \TeV$, using estimators $x_1 = \log_{10}(b_{LLP}),\  x_2 = n_{jet}$. $N_{samples}$ is the number of pseudo-data sets tested for the indicated number of events $N_{obs}$. Results for high-width are shown in brackets next to those for narrow-width. Note that a measurement of  $m_{parent}$ using LLP events is only possible for $\sim 1000$ observed decays for our benchmark point.}
	\label{tab:RESbfmassestimationresults}	
	\end{table}

	To evaluate the robustness of parameter estimation for this model under variation of the width of the heavy resonance, likelihood fits were performed for samples of narrow width events using template distributions from the high-width case, and vice versa. Using this method, $m_{LLP}/m_{parent}$ can be estimated with similar precision to that achieved using the correct template distributions, but with a small bias. The average $m_{LLP}/m_{parent}$ for narrow width samples under the high-width assumption is higher than the true value by approximately 1\%, but this is smaller than the spread of the distribution of best-fit masses up to $N_{obs}=1000$. The average mass ratio for high-width samples under the narrow width assumption is biased slightly lower than the true value, by approximately 3\%. The LLP to parent mass ratio can therefore be reliably extracted from the LLP boost distribution measured at MATHUSLA alone, once the production mode has been classified as Heavy Resonance, regardless of the resonance width.

	On the other hand, estimation of $m_{parent}$ using the incorrect width fails completely. In both cases, the best-fit masses were at the edge of the range of masses simulated, at least 50\% off of the true value, indicating that it is not possible to estimate $m_{parent}$ without an estimate or assumption of the heavy resonant particle's width, regardless of $N_{obs}$.  This emphasizes the importance of main detector resonance searches to determine all parameters of this LLP simplified production model.

In the high statistics limit, the width and mass of the resonance could be measured independently of a separate resonance search by performing a full  three-dimensional likelihood fit for the LLP mass, parent mass, and parent width. This is beyond our scope, but we demonstrate that this is possible in principle by classifying pseudo-data sets as low- or high-width based on the shape of the sample's boost distribution, using the ratio of maximum likelihoods under the two width hypotheses as a discriminant. The results are summarized in Table \ref{tab:RESwidthestimationresults} for true $m_{parent}=1\ \TeV$, $m_{LLP}=300\ \GeV$. It is clear that at least the two extremes of possible heavy resonance widths can be reliably distinguished, and that with enough observed events the heavy resonance's width could be estimated simultaneously with BSM particle masses.
  	\begin{table}
	\centering
	\begin{tabular}{|c|c|c|c|}
	\hline
	$N_{obs}$& 10 & 100 & 1000 \\
	$N_{samples}$ & 22812 (5104) & 2279 (508) & 228 (51) \\
	Classification accuracy & 0.81 (0.57) & 0.96 (0.96) & 1.0 (1.0) \\
	\hline
	\end{tabular}
	\caption{Heavy Resonance model width classification results from likelihood ratios of  $\log_{10}(b_{LLP})$ distribution with true $m_{parent}=1\ \TeV$, $m_{LLP}=300\ \GeV$, and $\Gamma_{parent}=10\ \GeV,\ 300\ \GeV$. Results for high-width are shown in brackets next to those for narrow width. $N_{samples}$ is the number of pseudo-data sets tested for the indicated number of events $N_{obs}$.}
	\label{tab:RESwidthestimationresults}	
	\end{table}

\section{Uncertainty in Matching LLP Decay to Production Vertex}	
\label{section:pileup}

The analyses above demonstrate that combining information from MATHUSLA and the main detector in the event of an LLP discovery would allow both the production mode and the mass parameters of the underlying simplified model to be determined. For simplicity, we have so far assumed that each LLP decay at MATHUSLA could be associated with a single hard primary vertex at the HL-LHC. However, this is unlikely to be the case,  partially due to some uncertainty in the LLP boost measurement at MATHUSLA, which narrows down the bunch crossings that produced the LLP down to a few, but more importantly because of the high pile-up multiplicity of 140-200 during the HL-LHC running conditions~\cite{ATL-UPGRADE-PUB-2013-014,Soyez:2018opl},

Even given the fact that `interesting' hard events are in absolute terms fairly rare in $pp$ collisions, and that the primary vertex with the highest associated total energy in the main detector is likely to be the one that corresponds to the LLP production event for most production modes, it seems apparent that in a realistic analysis, each LLP observation at MATHUSLA will correspond to a list of candidate primary vertices, instead of just one. We must therefore ask how our classification and parameter estimation analysis would be affected by this complication. We find that our methods are robust and should be adaptable to account for an ambiguity in the exact production vertex.

\subsection{Impact on Production Mode Classification}
If there is an `ensemble' of multiple candidate hard primary vertices per LLP, then instead of cuts on the fraction of events meeting certain criteria, the classifier would impose cuts on the fraction of \emph{ensembles} of candidate primary vertices meeting criteria. This could be as simple as demanding that at least one of the candidate vertices corresponding to an LLP observation satisfy certain conditions. We explore the impact of such a change for each of the simplified models. 
\begin{enumerate}
	\item  Since the requirement to classify a sample as Exotic $B$-meson Decay is a lack of hard scattering events, the additional candidate hard scatters will not lead to other models being classified as BB. However, it is likely that BB samples will be more likely to fail this criterion. The maximum fraction of events  across all ensembles with jets of $p_{T} > 20\ \GeV$ can be decreased to accommodate this, or we could impose a cut on not the fraction of events with a hard jet, but the fraction of ensembles where a certain fraction of events have a hard jet. This can be optimized in a realistic analysis. 
	\item  The correct production vertex in the Charged Current model can still be identified by a hard lepton, which does not occur often in the detector. Therefore, classification for this model should be relatively unaffected. 
	\item Classification of the Heavy Parent model relies on jet multiplicity, which is vulnerable to contamination from spurious vertices. This contamination can be reduced by raising the minimum $p_{T}$ and/or jet multiplicity for a candidate vertex to pass selection criteria. There may be a small penalty to classification accuracy for this model, but it should still perform similarly to what we demonstrated in this work.      
	\item The main signature used to identify Exotic Higgs Decay samples is the presence of well-separated jets from VBF. While correct identification of the production vertex may be more difficult for the GF production channel, the presence of VBF-like jets in roughly 10\% of the sets of candidate vertices should be roughly unaltered, as should the classifier's performance for this model. 
	\item The classification of the DPP model uses only LLP boost information, and is therefore unaffected by uncertainty in the production vertex identification. 
	\end{enumerate}
With only small modifications, the classification procedure outlined in Section \ref{section:classification} should therefore be remarkably robust under the introduction of multiple possible LLP production vertices. We turn now to a discussion of how parameter estimation would proceed in these circumstances.

\subsection{Impact on Parameter Estimation}
For the Exotic $B$-meson Decay, Exotic Higgs Decay, and Direct Pair Production models, parameter estimation only uses LLP boost, so it is unaffected. For the Charged Current and Heavy Parent models, the correct production vertex can likely be identified in a high fraction of events due to associated hard objects or leptons, so parameter estimation can proceed as normal. Finally, for the Heavy Resonance model, estimation of $m_{LLP}/m_{parent}$ uses only LLP boost and is unaffected. Estimation of $m_{parent}$ using the ISR spectrum is unreliable even when the correct vertex is identified, and is likely to rely on separate resonance searches performed with the main detectors. Therefore, the ability of MATHUSLA to estimate BSM particle masses is essentially unaffected by the complication of multiple possible LLP production vertices.

\section{Conclusions}
\label{section:conclusion}
Many well-motivated scenarios for BSM physics include long-lived particles, and it is important that the experimental search program for long-lived particles continue to be developed and expanded. The proposed MATHUSLA detector represents a vital part of that program, able to probe parameter space inaccessible by any other experiment. Allowing MATHUSLA to act as a Level-1 Trigger for CMS is required to take full advantage of both detectors' potential. This would enable an analysis of LLP observations at MATHUSLA to be augmented by observation of their production events in the main detector. With information from both experiments, both the production and decay topology of long-lived particles, as well as the BSM particle masses, can be accurately determined.

We have shown that for heavy LLPs, an accurate diagnosis at the simplified model level can be achieved $>90\%$ of the time with only 100 observed events, and $\approx 98\%$ of the time with 1000 observed events. Similar performance is possible for lighter LLPs, except for the Heavy Parent and Charged Current models with squeezed spectra $m_{LLP}\sim m_{parent} < 100\ \GeV$, where classification fails because the associated objects being used to identify the production mode become too soft. With similar statistics, the underlying parameters of the simplified model, like LLP and parent particle mass, can be measured with $\sim 10\%$ precision or better in most cases. The exception is the Heavy Resonance production mode, where the ratio of LLP to parent particle mass can be reliable extracted from the LLP boost distribution, but measurement of the parent particle mass would likely benefit from separate resonance searches conducted with the main detectors, in analogy to  mediator resonance searches in Dark Matter simplified models~\cite{Fairbairn:2016iuf,Chala:2015ama,Alexander:2016aln,Bagnaschi:2019djj}. 

This  performance is achieved with extremely simple cuts and analyses using robust, physically motivated features of LLP production events. Further work is sure to improve on our demonstrated classification accuracy and measurement precision. One case that will require further study is LLP production in exotic $B$-meson decays, where the masses of LLP decay products cannot be neglected, and the measurement of LLP boost at MATHUSLA is consequently more difficult. We have also discussed how to extend our work to include other LLP production scenarios like Heavy Parent decay to leptons + LLP, Heavy Parent with multi-step decay chains, or the consideration of a Higgs-like intermediate resonance of indeterminate mass. It is also imaginable that post-discovery, we suffer an embarrassment of riches, with many different LLP production and decay modes being observed. In that case, our analysis would have to be extended to allow for more than one dominant channel, but this work still demonstrates in principle how the information in each channel could be extracted. Finally, if the LLP lifetime is in the range $\bar b c \tau \sim 1 - 100 \ \meter$, the geometric distribution of DVs in MATHUSLA's decay volume could be used to extract the lifetime directly. Our methods are applicable not just to MATHUSLA and CMS, but also to other external LLP detector proposals, or even LLPs discovered using LHC main detectors alone. This emphasizes not only the great discovery potential of  new LHC detectors like MATHUSLA~\cite{Chou:2016lxi}, FASER~\cite{Feng:2017uoz} or CODEX-b~\cite{Gligorov:2017nwh}, but also shows that in the event of a discovery, however it took place, the origin of Long-Lived Particles can be uncovered in great detail. 

\vspace*{5mm}
\noindent \textbf{Acknowledgements:} 
We thank Erez Etzion, 
Henry Lubatti, Jessie Shelton, David Wandler, Steven Winter, and Charlie Young for helpful discussions.
The research of JB and DC was supported in part by a Discovery Grant from the Natural Sciences and Engineering Research Council of Canada, and by the Canada Research Chair program. The research of JB was supported in part by a Canada Graduate Scholarship from the Natural Sciences and Engineering Research Council of Canada. Computations were performed on the Niagara supercomputer at the SciNet HPC Consortium \cite{Niagara,SciNetLessons}. SciNet is funded by: the Canada Foundation for Innovation; the Government of Ontario; Ontario Research Fund - Research Excellence; and the University of Toronto. This research was enabled in part by support provided by Compute Canada (www.computecanada.ca).
\appendix

\section{Simulation Details}
	\label{section:appendix}
	This Section contains a detailed description of how LLP production events were simulated for each of the simplified models considered in this work. 
	\begin{enumerate}
	\item For the $B$-meson decay model, $b\bar{b}$ production was simulated with the MadGraph Standard Model. The $B$-mesons produced were identified in the simulation at hadronization level. If two were present, one was randomly chosen and manually decayed to two- or three-body final states consisting of SM electrons and/or mesons as well as the LLP in accordance with the relative branching ratios in~\cite{Evans:2017lvd,Bondarenko:2018ptm}, depending on the LLP type. Heavier $B$-baryons were present in approximately 10\% of events, but were ignored for simplicity. (A more realistic analysis would include them separately, but this would not affect our conclusions.) If a jet with $p_{T}>20\ \GeV$ corresponding to the $B$-meson chosen for LLP decay was present in the simulated event at detector level, it was removed from the event. A jet was defined as corresponding to a $B$-meson if the $\Delta R = \sqrt{(\Delta\phi^2) + (\Delta \eta)^2}$ between the jet and $B$-meson was less than $0.5$. This value was chosen by considering the distribution of $\Delta R$ to find the characteristic distance between $B$-mesons and their corresponding jets at detector level. The SM products of the exotic $B$ decay were not added back into the event. Since most $b\bar{b}$ production at the LHC is near threshold energy, and the $B$-mesons characteristically have momentum $\approx 5$ GeV, only $\approx 2\%$ of events contained jets with $p_{T}$ above the $20\ \GeV$ threshold at detector level. Therefore, the effect of completely discarding the SM products of the exotic decay is negligible. A reweighting of these events was performed by comparing the $B$-meson $p_{T}$ spectrum to one obtained from FONLL for $B$-mesons with $0<\eta<3$, roughly corresponding to those heading towards MATHUSLA~\cite{Cacciari:1998it}. 
	\item For the Charged Current model, the \textit{ChargeWN} MadGraph model from the LLP Simplified Model Library was used~\cite{Alimena:2019zri}. The simulated process was production of an (anti-)muon plus the lightest neutralino, mediated by an s-channel $W'^{\pm}$ boson. Because of the distinctive hard lepton present in these events, rendering analysis of jet variables irrelevant for model classification, jet matching was not performed for these samples. The choice was made to restrict to the case of decay of the $W'$ to muons plus LLPs. Decay to other leptons or lepton-universal decays would not present any additional challenges for LLP production mode identification.
	\item For the Heavy Parent model, the the minimal supersymmetric standard model (\textit{MSSM}) model was used~\cite{Nilles:1983ge,Haber:1984rc,Rosiek:1989rs,Skands:2003cj,Allanach:2008qq,Martin:1997ns}. In the one-jet decay mode the simulated process was squark anti-squark production, with each squark (anti-squark) decaying to either one quark or anti-quark as well as the lightest neutralino, which serves as the LLP. In the two-jet decay mode the simulated process was gluino pair production, with each gluino decaying to two quarks via an off-shell 10 TeV squark. Because the most important jets in these events originate from the hard process, jet matching was not performed on these samples. The choice was made to restrict the decay of the heavy parent particle to jets only, because this presents the most difficult case to distinguish from other simplified models. Decay to leptons plus LLPs would be identifiable by a large proportion of events containing multiple hard leptons. 
	\item For the Exotic Higgs Decay model, Higgs production was simulated using the \textit{heft} model in Madgraph \cite{PhysRevD.59.057504,PhysRevD.22.178,Kauffman1999}. The gluon fusion, vector boson fusion, and associated vector boson production modes were included. Jet matching was performed up to one jet for the gluon fusion channel, with the jet matching variable xqcut set to 15. Jet matching was performed to increase the reliability of simulated jet-related variables, since the only jets in these events are due to initial state radiation, and none are produced in the hard process. The Higgs boson was not permitted to decay. For each event, a pair of LLP 4-vectors with energy $m_{H}/2$ were manually generated pointing in a random direction and its opposite, as one would expect in Higgs decay to two scalars in the Higgs rest frame. These 4-vectors were then boosted into the lab frame using the Higgs momentum from simulation. 	
	\item For the Direct Pair Production model, the same \textit{MSSM} package was used, and pair production of the lightest neutralino as the LLP was the simulated process. The intermediate, t-/u-channel squark was set to $10\ \TeV$ to remove it as a dynamical degree of freedom. Jet matching was performed up to one jet, and the variable xqcut was set to $0.4 \times m_{LLP}$.  This choice of xqcut was made to produce smooth distributions of jet kinematics variables that were stable under small variation of xqcut. 
	\item 	For the Heavy Resonance model, the simplified model $Zp_{2LLP}$ was used \cite{Alimena:2019zri}. The process simulated was dark scalar pair production with a $Z'$ in the s-channel, with the dark scalar's width set to $1\times10^{-20}\ \GeV$ so that they did not decay in simulation. Jet matching was performed up to one jet, and the variable xqcut was set to $10 + 0.1\times m_{parent}$ after a similar optimization procedure to that used for Direct Pair Production.
	
	\end{enumerate}

\bibliographystyle{JHEP}
\bibliography{References}

\end{document}